\documentclass[oneside, a4paper, twocolumn, final]{article}
\usepackage[english]{babel}
\usepackage[utf8]{inputenc}
\usepackage[T1]{fontenc}

\usepackage{derivative}  % defines \odif{} \odv{} \pdv{}...
\usepackage{amsmath, amssymb, amsthm, mathtools, cancel}
\usepackage{graphicx, float, adjustbox, xcolor}
\usepackage[font=small,labelfont=bf,textfont=it]{caption}
\usepackage{hyperref}
    \hypersetup{
        colorlinks,
        linkcolor={red!50!black},
        citecolor={blue!50!black},
        urlcolor={blue!80!black}
    }
\usepackage{cleveref}
\crefname{chapter}{section}{sections}
\Crefname{chapter}{Section}{Sections}

\usepackage{booktabs}

\usepackage[left=2cm,right=2cm,top=3cm,bottom=3cm]{geometry}

    \newcommand{\half}{\frac{1}{2}} %make \frac{1}{2} with \half
    \newcommand{\define}{\equiv} %definition sign with \define
    
    \DeclareMathOperator*{\argmax}{arg max}

    \newcommand{\qq}[1]{\quad\text{#1}\quad}
    \newcommand{\vb}[1]{\boldsymbol{\mathrm{#1}}}

    \DeclarePairedDelimiter{\abs}{\lvert}{\rvert}

    \AtBeginDocument{%

    }

\NewDocumentCommand{\braket}{ o m m o }{%
    \ensuremath{%
        \IfValueT{#1}{{}_{#1}} % Prescript
        \langle #2 | #3 \rangle
        \IfValueT{#4}{_{#4}} % Subscript
    }%
}
\NewDocumentCommand{\bra}{ o m }{%
    \ensuremath{%
        \IfValueT{#1}{{}_{#1}} % Prescript
        \langle #2 |
    }%
}
\NewDocumentCommand{\ket}{ m o }{%
    \ensuremath{%
        | #1 \rangle
        \IfValueT{#2}{_{#2}} % Subscript
    }%
}
\usepackage{xparse}
\usepackage{ifdraft}
\usepackage{xcolor}

\DeclareDocumentCommand{\warn}{ m g }{%
    \ifdraft{%
        \colorbox{red}{\textcolor{white}{\strut\textbf{\MakeUppercase{\IfNoValueTF{#2}{WARNING!}{#2}}}}}%
    }{}%
    \PackageWarning{UserWarning}{#1}%
}

\usepackage{csquotes} % recomended by biblatex - don't know why
\usepackage[style=numeric-comp, autocite=inline, giveninits=true, sorting=none, url=false, isbn=false, maxbibnames=2]{biblatex}
    \DeclareNameAlias{labelname}{given-family}  % makes \citeauthor include given (first) name as well as family name
\usepackage{authblk}  % title and author stuff
\title{Fault-tolerant interfaces for modular quantum computing on diverse qubit platforms}

\author[1,2]{Frederik K. Marqversen}
\author[3,4]{Gefen Baranes}
\author[3,4]{Maxim Sirotin}
\author[3]{Johannes Borregaard}

\affil[1]{Department of Physics and Astronomy, Aarhus University, DK-8000 Aarhus C, Denmark}
\affil[2]{Kvantify ApS, DK-2300 Copenhagen S, Denmark.}
\affil[3]{Department of Physics, Harvard University, Cambridge, Massachusetts 02138, USA}
\affil[4]{Massachusetts Institute of Technology, 77 Massachusetts Avenue, Cambridge, MA 02138, USA}

\date{August 2025}

\begin{document}

\twocolumn[
  \begin{@twocolumnfalse}
    \maketitle
    \begin{abstract}
        \noindent\itshape
        Modular architectures offer a scalable path toward fault-tolerant quantum computing by interconnecting smaller quantum processing units (QPUs) provided that high-rate, fault-tolerant interfaces can be realized across modules. We present a comprehensive analysis and comparison of known and new methods for establishing such interfaces, including lattice surgery, transversal gates, and novel grow-and-distil protocols based on code growing and logical distillation. Using the surface code, we identify optimal interface strategies across a wide range of hardware parameters, such as gate fidelities, entangling rates, and memory resources, and estimate the requirements to achieve logical error rates of $10^{-6}$ and $10^{-12}$. Our results establish when the interface become a bottleneck in the computation and provide guidance for experimental implementations with superconducting, atomic, and solid-state hardware.
    \end{abstract}
    \vspace{1cm}
    \end{@twocolumnfalse}
]

\section{Introduction} \label{sed:introduction}

Quantum computing has made remarkable advancements in recent years, achieving systems with hundreds of physical qubits and demonstrating the first successful implementations of quantum error correction (QEC). This progress has been driven by various qubit platforms including atomic \cite{Egan2021,Bluvstein_2024_neutralatomQC,rodriguez2024,reichardt2024,RyanAnderson2024,Chen2024, bluvstein2025architectural} and superconducting qubits \cite{lacroix2024,Gupta2024,Putterman2025}. Despite these advances, the full potential of quantum computing remains out of reach. Many of the transformative applications envisioned for quantum computers, including breakthroughs in materials design \cite{Alexeev2024}, quantum chemistry \cite{McArdle2020}, and cryptography \cite{gidney2025factor}, still require capabilities far beyond those of current devices. State-of-the-art quantum algorithms demand millions of qubits and substantially lower error rates than currently demonstrated to outperform classical computing \cite{Lee2021,beverland2022,dalzell2023, gidney2025factor}. 

The complexity and cost of building a monolithic quantum computer capable of running fault-tolerant algorithms with millions of qubits can be prohibitive. A promising alternative is a modular quantum computer that interconnects smaller quantum processing units (QPUs) with independent control to simplify engineering and reduce costs. Computations across QPUs can then be performed by establishing qubit entanglement, enabling the transfer of quantum states and gate operations via quantum teleportation \cite{Monroe2014,sinclair_2025neutral_atoms_interconnect}.

To enable reliable computation on modular quantum computers, fault-tolerant operation is required not only within QPUs but also across their interfaces, necessitating the use of QEC. By encoding logical qubits in physical qubits, QEC allows fault-tolerant logical operations throughout the system. Recent studies have explored several strategies for establishing fault-tolerant interconnects between QPUs through the generation of logical Bell pairs, including  transversal gates \cite{stack2025}, lattice surgery \cite{ramette_fault-tolerant_2024}, and logical distillation \cite{pattison_fast_2024}. These approaches impose different requirements on inter-QPU entangling rates, the fidelity of physical Bell pairs, and the qubit overhead necessary to achieve fault tolerance. Given the diversity of quantum computing hardware and the wide variation in their physical parameters, it remains an open and timely question, which approaches are most promising for different qubit platforms.

\begin{figure*}
    \centering
    \includegraphics[width=\linewidth]{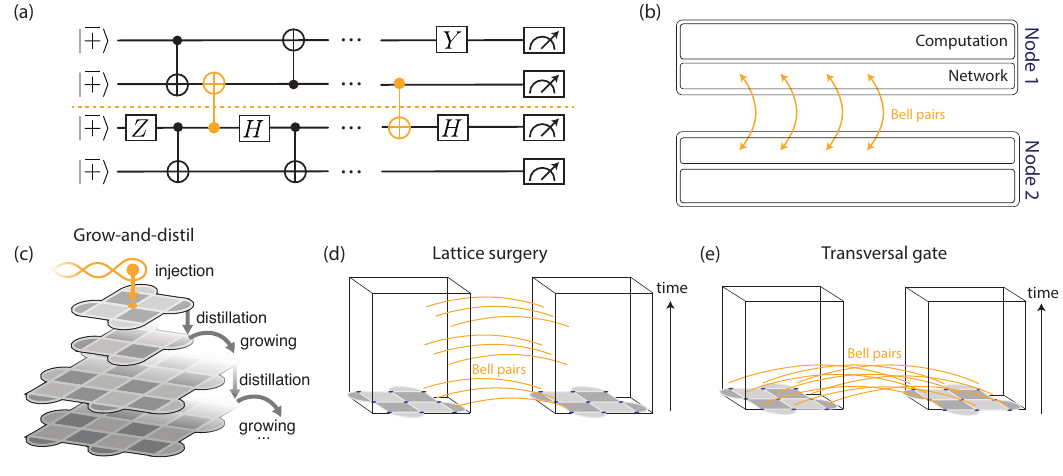}
    \caption{
        (a) Distributed algorithm between two nodes, requiring local logical gates (black) as well as distributed logical gates (orange), both with high fidelity. 
        (b) Physical implementation of two nodes, each with a computation zone and a network zone, with physical Bell pairs distributed between both nodes. We are assuming there is a direct quantum connection between qubits in network and computation zones.
        Three main approaches for fault-tolerant distributed QC are presented in (c-d). 
        % (c) Logical distillation with a sequence of QED codes interleaved with growing, as described in ref. \cite{pattison_fast_2024}.
        (c) Injection and logical distillation interleaved with growing (grow-and-distil).        
        (d) Lattice surgery
        (e) Transversal gate.
    }
    \label{fig:distributed QC concept}
\end{figure*}

In this work, we address this question by introducing efficient grow-and-distil protocols for logical distillation using code-growing techniques~\cite{li_magic_2015} and providing a comprehensive comparison with other known methods for establishing fault-tolerant interfaces. Considering the surface code \cite{fowler2012surface}, we perform a cross-method optimisation that incorporates combinations of physical Bell pair distillation, lattice surgery, transversal entangling gates, logical distillation, and code-growing to determine which approaches maximise the logical Bell pair generation rate across a broad range of hardware parameters. These parameters include qubit decoherence times, local gate speeds and fidelities, network entangling rates and error rates, and the qubit resources available within QPUs. Through this analysis, we identify the operational regimes in which each method is optimal, providing a reference for experimental implementations of scalable, fault-tolerant modular quantum computing across diverse platforms, including superconducting \cite{majidy2024building, google_2024_below_threshold_SC}, atomic \cite{sinclair_2025neutral_atoms_interconnect, bluvstein2025architectural}, and solid-state qubits \cite{bartling2024universal, knaut_entanglement_2024}. 

\section{Grow-and-Distil}

We consider a distributed quantum computing architecture composed of multiple nodes, each containing a network zone and a computation zone, as illustrated in \cref{fig:distributed QC concept} (a–b).
The network zone is dedicated to generating physical Bell pairs and enabling fault-tolerant distributed logical gates, while the computation zone executes the fault-tolerant quantum algorithm.
\Cref{fig:distributed QC concept} (c-e) showcases the three main approaches for distributed fault-tolerant logical gates on the surface code: logical distillation, lattice surgery, or transversal gates with physical Bell pairs \cite{pattison_fast_2024,sinclair_2025neutral_atoms_interconnect,ramette_fault-tolerant_2024}.

Previous work on logical distillation~\cite{pattison_fast_2024} showed that it is possible to  generate high-fidelity logical Bell pairs from only a few noisy physical Bell pairs. First, state injection techniques are employed to inject noisy physical Bell pairs into logical qubits.
Then, quantum error detection (QED) at the logical level is used to distil high-fidelity logical Bell pairs via fault-tolerant Clifford operations and measurements.
This approach allows for entanglement distillation with a constant encoding rate and supports fault-tolerant logical gates using only a small number of noisy physical Bell pairs.  However, it  demands a large memory buffer for the logical encoding at each node.

To reduce the memory requirements, we introduce a novel grow-and-distil method where stages of code growing are interspersed with distillation steps. This allows for the early distillation stages to be performed on small noisy logical qubits. The reduction in logical qubit size directly results in a reduction in space and the effect is amplified by the fact that earlier stages of distillation have much higher throughput and more aggressive post-selection. Slightly reducing the size of earlier stages thus reduces the total qubit overhead significantly. Due to the smaller logical code sizes leading to larger logical error rates, it is conceivable that logical gate errors would play a more significant role than assumed in ref. \cite{pattison_fast_2024}. However, we find that the logical error rates do not affect the distilled error rates much as long as they are less than Bell state errors. 

%An additional consequence of allowing for code growing is that the space of possible distillation sequences increases considerably. Efficiently searching for the optimal sequence is thus much more involved. Details on how this is done are included in \cref{app:sequence optimisation}.

Formally, we describe a grow-and-distil sequence $S$ as consisting of a sequence of stages $(S_i)$. The very first stage is always state injection, which injects the physical Bell pair into a small surface code. Apart from the first stage, stages can be of two types: Logical distillation and code growing. Logical distillation is done as in ref. \cite{pattison_fast_2024} using any error correcting code $[[n, k, d]]$. In this work, we include the same code-set as was used in that original work, which includes 512 distinct codes up to size $n=30$. Code growing is performed as described in \cite{li_magic_2015}. Taking as input a surface code logical qubit, two code patches are added by lattice surgery to finally output a surface code of a bigger size. We estimate that the logical error rate from code growing is approximately twice that of local logical gates on the initial code distance. Details on the procedure and numerical validation of this model are discussed in \cref{app:growing}.

\subsection{Error bounds}
We use analytical bounds to evaluate the performance of the grow-and-distil sequence. As such, we provide lower performance bounds rather than approximate optimal values. The relevant bounds for this work are similar to those introduced in ref. \cite{pattison_fast_2024}, but we have to explicitly take into account the logical gate errors due to the smaller sized logical codes.

Consider a distillation stage using general error correcting code $[[n, k, d]]$. Using the general parallelised unencoding circuit presented in the appendix of \cite{pattison_fast_2024} we know that unencoding can be performed by a depth $D$ circuit where
\begin{equation}
    D = 3n - 2 - k.
\end{equation}
Note that this circuit is applicable to a general error correcting code. By specialising the unencoding for each code type, circuits of much smaller depth can be achieved. 

Let the surface code size of the logical qubits be $L$ with logical gate error rate $p_L$. Given input states with error rates $p_\mathrm{in}$, the probability that the output of the distillation stage must be discarded $p_\mathrm{fail}$ is bounded by the probability that all the inputs have no errors and that no errors happened during the depth $D$ unencoding:
\begin{equation}
    1 - p_\mathrm{fail} \geq (1 - p_\mathrm{in})^n (1 - p_L)^{nD} \define (1-q)^n.
\end{equation}
Here we define the error rate $q$, which is the probability that any specific qubit after unencoding has an error.

The output from a distillation stage after post-selection can include a logical error only if that error was not detected. The probability of such an error $p_\mathrm{out}$ is thus equal to the probability of the occurrence of an undetectable error. For a distance $d$ code this requires at least $d$ individual errors. The probability that no error was detected will be at least the probability of no errors occurring. This gives the following bound on $p_\mathrm{out}$:
\begin{equation}
    p_\mathrm{out} \leq \frac{\Pr(\abs{E} \geq d)}{\Pr(\abs{E} = 0)}
        =
    \frac{1 - \sum_{i=0}^{d-1} \Pr(\abs{E} = i)}{\Pr(\abs{E} = 0)},
\end{equation}
where $\abs{E}$ is the number of errors after unencoding but before post selection. A final bound is obtained by doing the substitution
\begin{equation}
    \Pr(\abs{E} = i) \rightarrow \binom{n}{i} q^i (1 - q)^{n - i}.
\end{equation}

By comparison with the bounds presented in ref. \cite{pattison_fast_2024}, it is evident that the bounds presented here also can be obtained by the direct substitution $p_\mathrm{in} \to q$. The error rate $q$ can thus be interpreted as an effective error rate that takes into account all of the effects of local logical gate errors.

The error rate of logical gates $p_L$ performed between surface code encoded qubits of size $L$ is modelled as
\begin{equation}
    p_L \propto \left( \frac{p_b}{p_b^*} \right)^{\mathmakebox[0pt]{\frac{L}{2}}},
\end{equation}
where $p_b$ is the error rate of local gates on physical qubits. For the values of the bulk threshold $p_b^*$ and proportionality constant, we take the numerically estimated values presented in \cite[Suppl.]{ramette_fault-tolerant_2024}.

\subsection{Distillation throughput}
Given a distillation sequence $S$ and an amount of space/memory $M$ that is allocated for networking, we wish to determine the expected rate at which successfully distilled Bell pairs can be produced. There are two regimes to consider, input limited and memory limited.

Consider initially a case where memory is not the limiting factor $M \to \infty$. The optimal pipeline strategy is to fully parallelise the distillation process, meaning that we allow multiple unencoding circuits for each stage to be running simultaneously. We refer to each actively running circuit as a stage instance. In practice, this is simply achieved by immediately initialising instances when enough inputs are available. This is called a balanced pipeline since instances never have to be actively paused to not exceed the memory constraint. In steady state, the outputs will be distilled at a rate
\begin{equation}
    r^\mathrm{out}_S = E_S r_\mathrm{bell},
\end{equation}
where $E_S$ is the \emph{encoding rate} of sequence $S$. The encoding rate is equal to the expected number of successfully distilled logical Bell pairs per input physical Bell pair. The encoding rate depends only on the error rate of the input physical Bell pairs and on the local physical gate error rate. Thus, for a given hardware, $E_S$ is an intrinsic quality of the given sequence $S$.

It turns out that for many pipeline problems, the above strategy can be applied even in memory limited cases. The balanced pipeline will take up some amount of memory, and the exact amount will depend both on the details of $S$ and the input rate $r_\mathrm{bell}$. Given a memory constraint $M$, the idea is to artificially reduce the input rate $r_\mathrm{bell}$ such that the balanced pipeline satisfies the constraint. In practice, this means that there exists an upper \emph{input rate cap} $C_S$ restricting the rate at which inputs can be supplied.
\begin{equation}
    r^{\mathrm{in}}_S = \min(r_\mathrm{bell}, C_S).
\end{equation}
The value $C_S$ is intrinsic to the sequence $S$ in an equivalent fashion to $E_S$. 

Using this $r^{\mathrm{in}}_S$ the pipeline is balanced in both the input limited and memory limited cases, and so, outputs will be produced at a rate
\begin{equation}
    r^\mathrm{out}_S = E_S r^{\mathrm{in}}_S = E_S \min(r_\mathrm{bell}, C_S).
\end{equation}
The implications are that for a given sequence $S$ and fixed memory constraint $M$, we can expect the distillation rate to increase linearly with the rate at which physical Bell pairs can be prepared. This will hold only to a point $r^{\mathrm{in}}_S \leq C_S$, above which inputs have to be discarded or redistributed due to insufficient space, and the distillation rate plateaus.

How $E_S$ and $C_S$ are computed for a given sequence $S$ along with other general details on this subject are presented in \cref{app:distillation rate}.

\subsection{Growing vs. no growing}
\begin{figure}
    \centering
    \includegraphics[width=\linewidth]{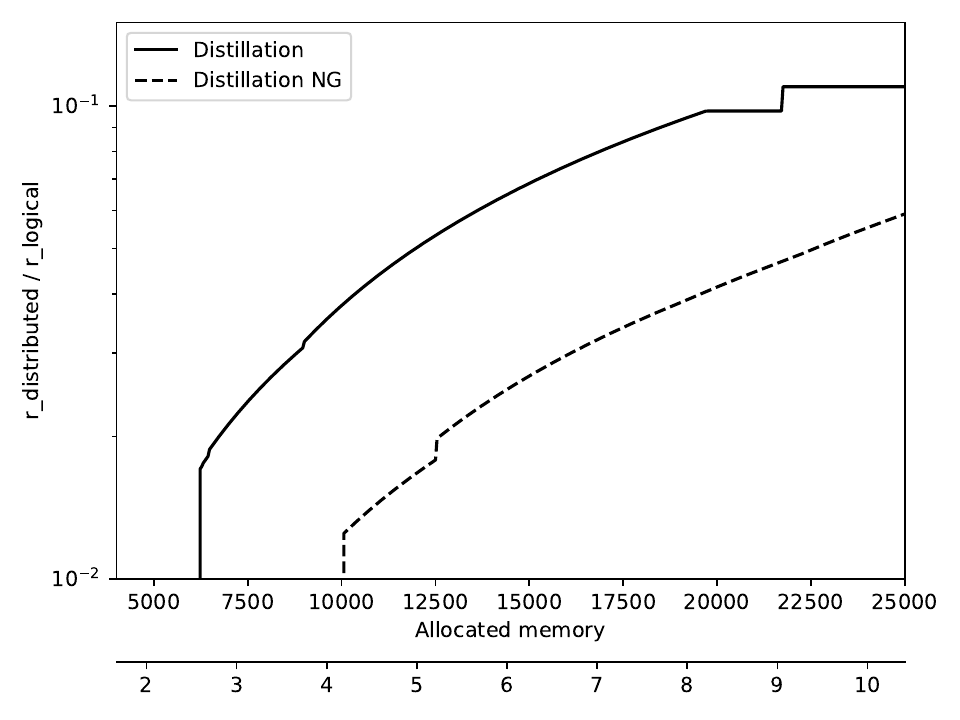}
    \caption{Rate of distributed logical Bell pairs $r_\mathrm{distributed}$ as a function of local memory dedicated to distillation. The two $x$-axes represent the amount of allocated space for distillation in terms of physical qubits (upper) and logical qubits (lower) respectively. Includes rates from both distillation with code growing and distillation without code growing (NG) with a target Bell pair error rate of $10^{-12}$. The figure is made assuming physical Bell pairs are produced at a rate equal to the time of local physical gates. Other parameters are initial physical Bell pair error rate of 1\%, physical gate error rate of 0.1\%, and idling errors per physical gate of $10^{-6}$. }
    \label{fig:distillation comparison}
\end{figure}

In \cref{fig:distillation comparison} we compare the performance of distillation with code growing to distillation without code growing for a target Bell pair error rate $10^{-12}$. The comparison is based on a physical Bell pair rate equal to the local physical gate rate, initial physical Bell pair error rate of 1\%, physical gate error rate of 0.1\%. Sequences without growing are evaluated by direct injection into a distance 3 surface code followed by a growing stage, immediately growing the code to the target logical qubit size.

Both of the curves in \cref{fig:distillation comparison} experiences a lower cutoff point, below which the available space is insufficient to run distillation. Also, both curves exhibits kinks. This is because both curves are the result of multiple distinct distillation sequences. The gradual increases in distillation rate are due to more parallelisation of a fixed distillation sequence. The kinks, on the other hand are the result of a new optimal sequence becoming available. Finally, the plateaus observed for the upper curve are the results of hitting the limit on the input rate for the currently best sequence. Note, that the final plateau is not actually a fundamental upper limit on the distillation rate. Provided more memory, a new sequence with a higher encoding rate might become available which then will allow further increasing the distillation rate, resulting in a second step in the curve.

From \cref{fig:distillation comparison} we observe that distillation with growing is possible at space $\geq 6,230$ physical or equivalently $\geq 5.16$ logical qubits, much lower compared to distillation without code growing possible at space $\geq 10,000$ physical or $\geq 8$ logical qubits. Also, we find that by including growing, networking rate is improved by at least a factor of 1.85 across \cref{fig:distillation comparison}.

\section{Direct gate methods}
Alternatives for implementing non-local connections is implementing direct distributed gates with the noisy Bell pairs, either using lattice surgery or transversal gates, as described in \cite{sinclair_2025neutral_atoms_interconnect,ramette_fault-tolerant_2024} and illustrated in \cref{fig:distributed QC concept}.
In the lattice surgery approach, two surface-code patches are merged across a shared boundary using a set of stabiliser measurements that span the seam.
These inter-module parity checks are performed using noisy Bell pairs and additional local operations, effectively stitching the patches into a larger code during the logical operation.
Alternatively, transversal gate schemes apply logical gates by teleporting physical-level operations across aligned surface-code patches using distributed Bell pairs.
Each qubit in one logical patch is paired with a corresponding qubit in the other, and a teleported entangling gate is performed transversally using a network of physical Bell pairs.
Recent results \cite{ramette_fault-tolerant_2024} show that fault-tolerant logical gates can be implemented in this way even when the Bell pairs exhibit error thresholds as high as 10\%, provided that the local gates remain below the bulk surface-code threshold (of $1\%$ for the surface code).
This method leverages the intrinsic robustness of the QEC codes to boundary noise.

\subsection{Logical Bell error}
To get a fair comparison between all three models considered in this work, we utilise the entire amount of memory that is allocated when modelling the throughput of the transversal and lattice surgery procedures. This is done by parallelising the procedures as much as possible within the constraints of the given space and physical Bell pair rate. Furthermore, the execution time is important for the final throughput. Transversal gates require only a single round of syndrome extraction by utilising correlated decoding \cite{zhou_algorithmic_2024,cain_correlated_2024}. Lattice surgery, on the other hand, requires the full $L$ rounds for an $L \times L$ surface code since the values of the measured syndromes within the seam between qubits are used for doing logic \cite{fowler2012surface}. Together with qubit idling this is a significant difference. A transversal gate requires a total of $L^2$ physical Bell pairs before it can be executed, whereas lattice surgery can be done in $L$ steps of $L$ Bell pairs. This means that the Bell pairs for lattice surgery are idling less than those for transversal gates, leading to smaller errors along the seam.

To model logical errors introduced by the direct distributed gate procedures, we use \cite{ramette_fault-tolerant_2024}:
\begin{gather}
    p_L
    \propto \left( \frac{p_s}{p_s^*} \right)^{\mathmakebox[0pt]{\frac{L}{2}}} 
    + \left( \frac{p_b}{p_b^*} \right)^{\mathmakebox[0pt]{\frac{L}{2}}} 
    + \sum_{j = 1}^L \left( \frac{p_s}{p_{1s}^*} \right)^{\mathmakebox[0pt]{\frac{j}{2}}} \left( \frac{p_b}{p_b^*} \right)^{\mathmakebox[0pt]{\frac{L-j}{2}}}
        \intertext{with}
    p_{1s}^* = p_s^* \left( 1 + \alpha_C p_b \frac{\sqrt{p_s^*}}{1 - \sqrt{p_b/p_b^*}} \right)^{-2}.
\end{gather}
\iffalse
Values: $p_b^* = 10.4\%$, $p_s^* = 0.75\%$, $\alpha_C = 1.4$, $c_b = 8.00\%$, $c_s = 15.43\%$, and $c_{sb} = 1.04\%$ \cite{ramette_fault-tolerant_2024}.
\fi
Here $p_b$ is the local physical gate error rate and $p_s$ the error rate of the distributed physical Bell pairs including idling errors. For the bulk threshold $p_b^*$, seam threshold $p_s^*$, and $\alpha_C$ we use numerically determined values from \cite[Suppl.]{ramette_fault-tolerant_2024}. Logical errors introduced by \emph{local} logical operations during distillation simply correspond to the special case $p_s=0$.

If the network is noisy, the local logical qubit size must be large enough to protect against the elevated error rates along the seam. In general, the direct distributed gate procedures require the local nodes to operate on qubits that are larger than required merely by local constraints. In our analyses, we find the smallest code size which can accommodate the sought after logical distributed error rate.

\subsection{Idling errors}
When comparing the performance of our distillation protocol to distributed transversal gates and lattice surgery, we must consider the fact that for the direct gate methods, a large number of physical Bell pairs are needed before the operation can be executed. The time it takes to prepare these will in many regimes be significant compared to idling error rates, and so, these effects must be accounted for. The same can be said in regard to the distillation of physical qubits. We model idling errors as a single-qubit depolarisation channel with error probability equal to that of the entire idling duration across all input qubits. Some qubits will be idling for less, all the way down to no idling for the final input qubit. Thus, this model is an overestimate of the actual effects of idling.

For logical qubit distillation, on the other hand, Bell pairs are injected into surface codes as soon as they arrive, offering protection against such idling errors. For later stages of a distillation sequence, where idling will be significantly larger than for earlier stages, the code size also will be larger. We thus model logical distillation without taking into account errors due to idling.

\section{Hardware agnostic method comparison}
\begin{figure}
    \centering
    \includegraphics[width=\linewidth]{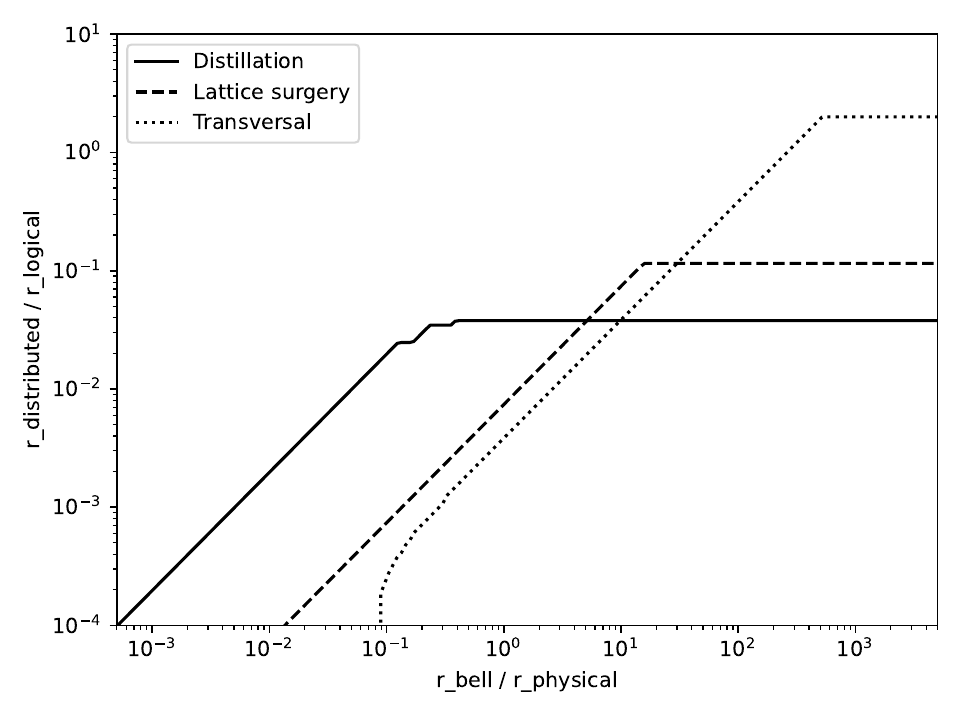}
    \caption{Rate of distributed logical Bell pairs $r_\mathrm{distributed}$ as a function of physical Bell pair rate $r_\mathrm{bell}$. These are given in units of local logical gate rates and physical gate rates respectively. Rates from each of the three methods: Distillation, logical CNOT by lattice surgery, and transversal logical CNOT are included. The figure is for initial physical Bell pair error rate of 1\%, a target Bell pair error rate $10^{-12}$, with allocated memory of 10\,000 physical qubits, physical gate error rate of 0.1\%, and idling errors per physical gate of $10^{-6}$.}
    \label{fig:method comparison}
\end{figure}

To describe the interplay between the different methods we include \cref{fig:method comparison} which shows, for each of the three methods, the rate at which logical Bell pairs are produced $r_\mathrm{distributed}$, equivalently the rate that logical distributed gates can be performed. The figure is made assuming an initial physical Bell pair error rate of 1\%, a target Bell pair error rate $10^{-12}$, with allocated memory of 10,000 physical qubits, physical gate error rate of 0.1\%, and idling errors per physical gate of $10^{-6}$.

The distillation curve has three different regimes. From the left: A linear ramp, a step, and then a constant plateau. For a particular amount of space, there will be some sequence with the largest encoding rate. This sequence will always be optimal so long as the sequence is not memory limited. This is the reason for the initial linear ramp extending to the left. Similarly, there will be some sequence with the largest distillation rate when memory limited. This means that we can talk about input and memory limited distillation in general, not just of individual sequences. These observations are properly formalised in \cref{app:sequence optimisation}. The step is the result of a single sequence which is optimal in the ``in between'' regime. Here we have only a singular step, although in general there could be multiple.

The curves for lattice surgery and transversal gate in \cref{fig:method comparison} are very similar. They both plateau because of the memory constraint. Lattice surgery plateaus earlier due to the required $L$ rounds of syndrome extraction, which leads to a long execution time. To the left of the plateau both curves first ramp down linearly, but then tapers off before finally reaching a point where they drop to zero. The tapering is simply a result of increasing logical qubit size required to reach the target Bell pair error rate of $10^{-12}$. The final cutoff point, left of which the target error rate cannot be reached, is the point where the input rate is so low that the accumulated idling errors bring the initial physical Bell pair error rate above the threshold of the surface code. Lattice surgery has a lower cutoff point due to the fact that idling is only accumulating in the time it takes for $L$ physical qubits to arrive, whereas for transversal gates the qubits are idling for the time of the full $L^2$ qubits.

\section{Analysis across hardware platforms} \label{sec:platforms}
Several physical platforms are under active development for distributed quantum computation, combining local quantum logic with photonic entanglement generation across remote nodes.
Among them, we consider three representative architectures: (1) neutral atoms in optical tweezers, (2) group IV solid-state defects in nanophotonic cavities, and (3) superconducting qubits with microwave links. These platforms differ significantly in connectivity models, gate mechanisms, communication fidelities, and memory scaling.
Here, we summarise the key assumptions made for each platform in our resource-performance analysis; full benchmarking data and references are detailed in \cref{app:exp_params}.

\paragraph{Neutral Atoms.}
This architecture uses individual neutral atoms (e.g., Rb or Yb) trapped in optical tweezer arrays, with laser-driven Rydberg interactions enabling high-fidelity local gates. Each node comprises thousands of atoms partitioned into computation and communication zones, supporting all-to-all connectivity via atom movement. Current demonstrations achieve local gate errors below 0.5\% and memory sizes up to 6,000 qubits with continuous reloading abilities \cite{Bluvstein_2024_neutralatomQC,manetsch2024tweezer,chiu2025continuous, bluvstein2025architectural}. 
% Gate times are limited by motion (~200\,$\mu$s), with future efforts targeting 10\,$\mu$s operation.
Distributed entanglement is generated using one of three photonic interconnects: (i) micro-cavities, (ii) a shared cavity, or (iii) free-space emission. Each method exhibits distinct scaling behavior in the required number of communication qubits versus Bell-pair generation rate \cite{sinclair_2025neutral_atoms_interconnect}.
We assume a total memory budget of $10^4$–$10^5$ atoms per node, and a normalised Bell-pair rate $r_{\mathrm{bell}} / r_{\mathrm{physical}}$ ranging from $10^{-1}$ to nearly $10^{3}$ depending on interconnect design, making this platform highly versatile for testing multiple QEC regimes.
In \cref{app:atoms landscape} we model the trade-offs between Bell pair rates and memory requirements using the analytical expressions from ref. \cite{sinclair_2025neutral_atoms_interconnect}.

\paragraph{Group IV Defects in Nanophotonic Cavities.}
Each node in this architecture contains a diamond-hosted group IV defect (e.g., SiV) coupled to a nano-photonic cavity \cite{stas2022robust,Cheng_SiV_BQC,bartling2024universal}, demonstrating large-distance entanglement distribution and blind quantum computation protocols \cite{knaut_entanglement_2024,baranes2025designing,Cheng_SiV_BQC}.
Communication is mediated by electron spins coupled to photons, while memory is stored in long-lived nuclear spins. 
Local logic and remote entanglement are performed via optically heralded spin-photon gates.
This architecture enables photonic interconnects for both local and distributed gates, but currently suffers from limited memory. 
We assume a memory size of 100–5\,000 qubits per node and a distributed-to-local gate rate ratio $r_{\mathrm{bell}} / r_{\mathrm{local}}$ between 0.1 and 1, consistent with future projections \cite{baranes2025designing,knaut_entanglement_2024}. 
% Coherence times are currently on the order of 1 second and expected to reach 10 seconds in future experiments, supporting low idling error rates and enabling error-corrected logic across photonic networks.

\paragraph{Superconducting Qubits.}
Superconducting quantum processors offer fast and high-fidelity local gates between qubits on the same chip \cite{google_2024_below_threshold_SC,acharya_suppressing_2023}. Each node consists of one or more cryogenic chips, with inter-chip and inter-node connections possibly established via microwave links \cite{majidy2024building,Wallraff_SC_entanglement_BellTest}. 
Unlike the other platforms, this architecture relies on lattice surgery for implementing logical operations, requiring $O(d)$ rounds of stabiliser measurements per gate \cite{zhou_algorithmic_2024,baranes2025leveraging, cain_correlated_2024,fowler2012surface}. This excludes the direct transversal gate method for the architecture.
We assume a future memory capacity of 1\,000–10\,000 qubits per node, and a distributed-to-local gate rate ratio $r_{\mathrm{bell}} / r_{\mathrm{local}}$ between $10^{-6}$ and $10^{-3}$.

In our cross-platform analysis, hardware platforms are numerically characterised by the following set of five parameters:
\begin{itemize} 
    \item $p_\mathrm{physical}$: Error rate of a single gate performed on physical qubits.
    \item $p_\mathrm{bell}$: Error rate of distributed physical Bell pairs
    \item $r_\mathrm{bell} / r_\mathrm{physical}$: Maximum rate of distributed physical Bell pairs in units of the physical gate rate
    \item $p_\mathrm{idle}$: Idling errors per one physical gate time
    \item $M$: Number of physical qubits (memory) allocated for networking
\end{itemize}
Together with a target logical Bell pair error rate $p_\mathrm{target}$, these parameters define a space in which the three methods: distillation, lattice surgery, and transversal gates, can be evaluated. We aim to estimate where in this space the platforms described in \cref{sec:platforms} are likely to operate, and to describe the qualitative characteristics of each region.

\subsection{Distributed gate rate landscape}
\begin{figure*}
    \centering
    \includegraphics[width=\linewidth]{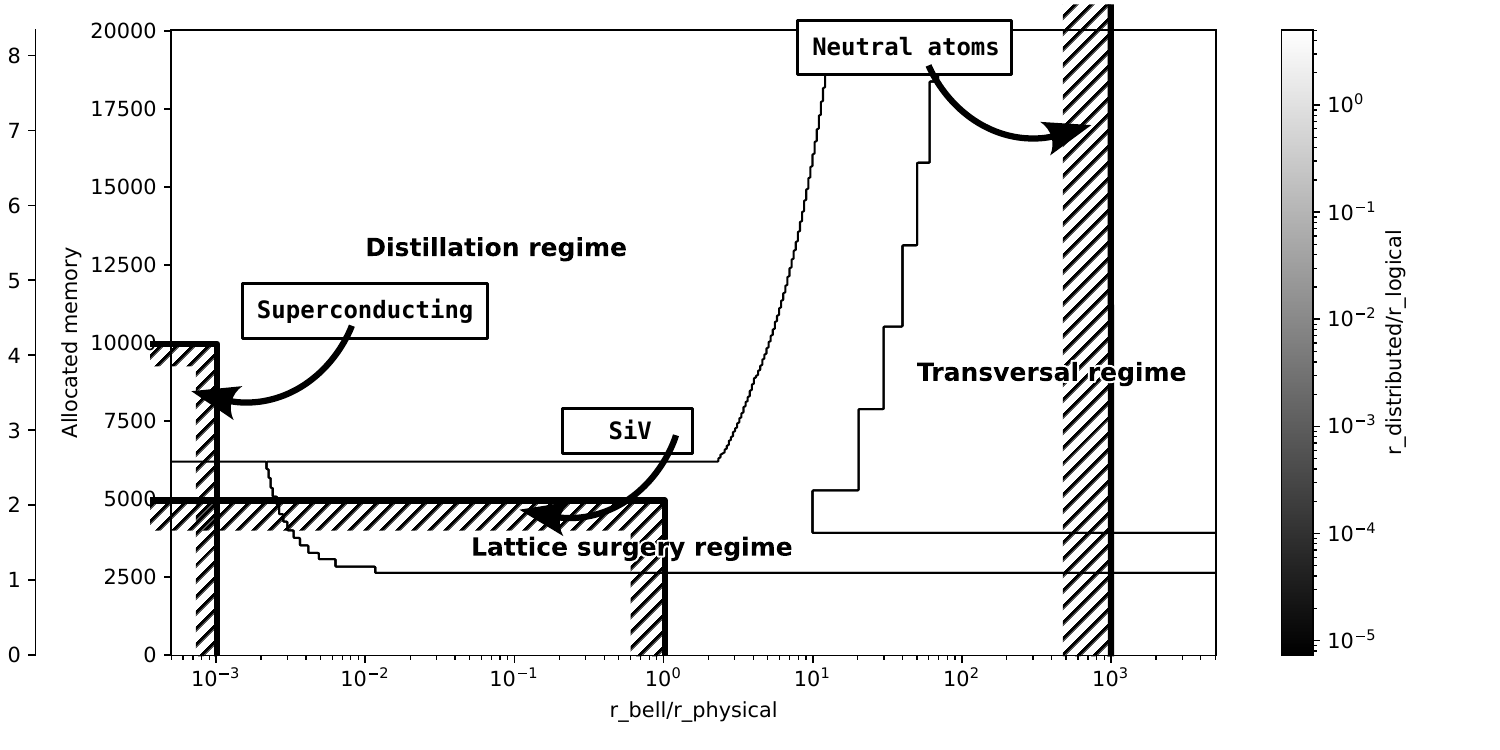}
    \includegraphics[width=\linewidth]{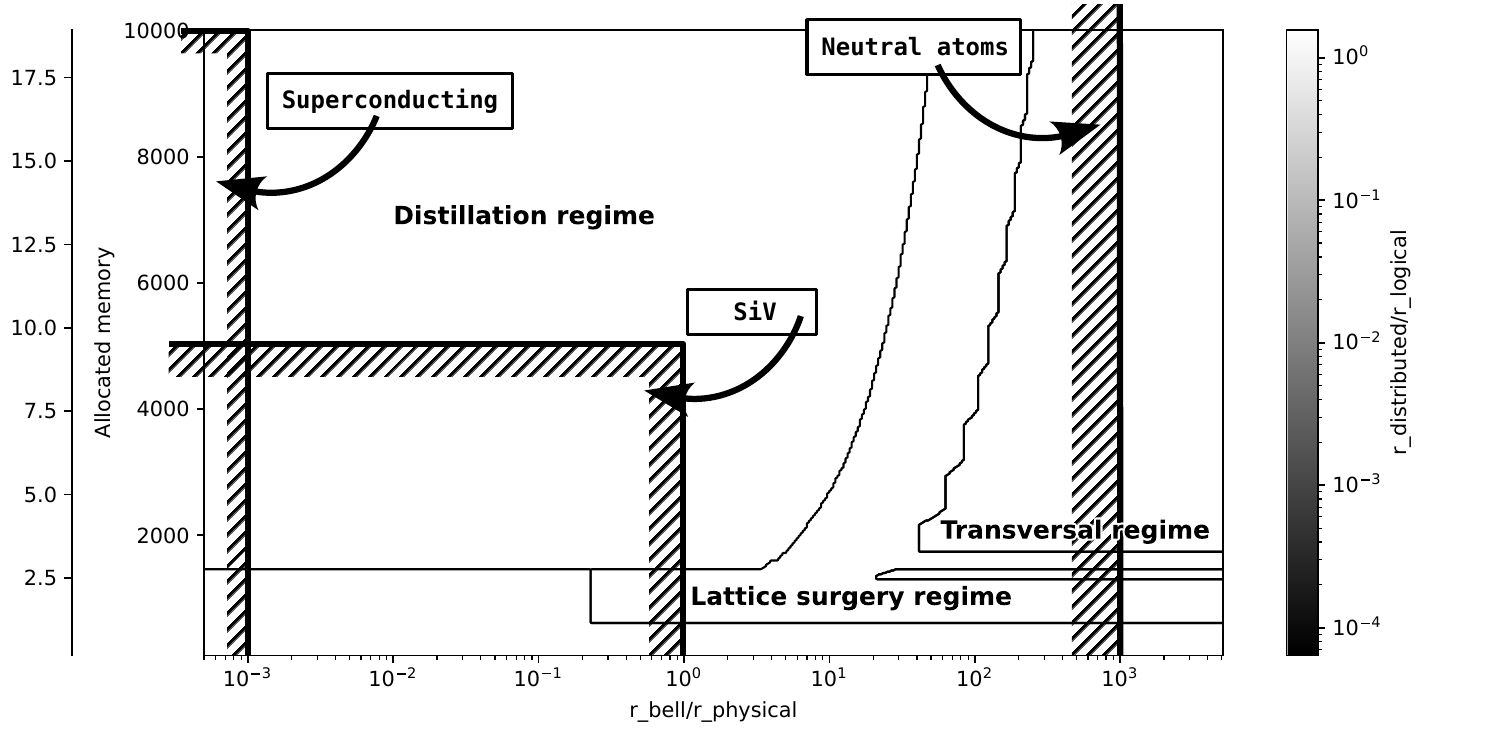}
    \caption{Logical Bell pair distribution rate $r_{\mathrm{distributed}}$ as a function of networking memory and physical Bell pair rate $r_{\mathrm{bell}}$. The two plots correspond to (top) $p_{\mathrm{bell}} = 1\%$, $p_{\mathrm{target}} = 10^{-12}$ and (bottom) $p_{\mathrm{bell}} = 5\%$, $p_{\mathrm{target}} = 10^{-6}$. Rates are expressed in units of the local physical gate rate $r_{\mathrm{physical}}$ and local syndrome extraction rate $r_{\mathrm{logical}}$. The $y$-axes show both physical (right) and logical (left) qubit counts. Each point reflects the highest rate achievable among the three methods: distillation, lattice surgery, and transversal gates. Black contour lines divide regions with different optimal methods. The full-white regions extending from the bottom left represent parameter regimes where no method achieves the target error rate. Discrete jumps are due to the discrete nature of surface code sizes and distillation protocols, and are not artifacts of resolution. Expected operational regions for neutral atoms, SiV, and superconducting qubits are outlined, each extending left and downward from the indicated boundaries.}
    \label{fig:top view}
\end{figure*}

In \cref{fig:top view}, we present two 3D landscapes in which the dependent variable (shown on the third colour axis) is the distribution rate of logical Bell pairs in units of the local logical gate rate, $r_{\mathrm{distributed}} / r_{\mathrm{logical}}$. The two free parameters are the physical Bell pair rate $r_{\mathrm{bell}} / r_{\mathrm{physical}}$ and the memory allocated for networking. We fix two of the remaining parameters at reasonable future hardware values: $p_{\mathrm{physical}} = 0.1\%$ and $p_{\mathrm{idle}} = 10^{-6}$ (see \cref{app:exp_params}). The remaining parameters, $p_{\mathrm{bell}}$ and $p_{\mathrm{target}}$, differ between the two plots. The top plot reflects an optimistic milestone with $p_{\mathrm{bell}} = 1\%$ and $p_{\mathrm{target}} = 10^{-12}$, while the bottom plot assumes a more conservative scenario with $p_{\mathrm{bell}} = 5\%$ and $p_{\mathrm{target}} = 10^{-6}$.

At each point in \cref{fig:top view}, all three networking methods are evaluated, and the maximum achievable distribution rate is plotted. This defines distinct regions where one method outperforms the others. Each platform's expected operational region as defined in \cref{sec:platforms} is also indicated. 

The reported networking rate assumes that the entire allocated memory is used for executing one of the three methods. This memory is therefore not available for local computation. Notably, we do not account for additional memory required to achieve a given $r_\mathrm{bell}$ through e.g. multiplexing. For a more detailed analysis including these effects for neutral atoms, see \cref{app:atoms landscape}.

For these figures, unencoding circuits are assumed to be implemented transversally, followed by a single round of syndrome extraction \cite{zhou_algorithmic_2024,cain_correlated_2024}. As noted in \cref{sec:platforms}, platforms lacking all-to-all connectivity (e.g., superconducting qubits) cannot perform transversal gates, leading to the time of performing unencoding being longer. However, our analysis including these effects shows that this has but negligible effect on distillation rates, since the dominant factor is discard rates, which remain unaffected. By defining $r_{\mathrm{logical}}$ as the rate of a single round of local syndrome extraction (rather than logical gate rate), we enable a consistent comparison across all three platforms in \cref{fig:top view}.

To achieve meaningful rates using direct-gate methods under the bottom-plot scenario in \cref{fig:top view}, we incorporate physical distillation. Here, the initial 5\% error Bell pairs are distilled to 1\% before being used by higher-level protocols. The memory cost of this distillation process is fully accounted for in the plot.

Physical distillation is analysed similarly to logical distillation, with the important distinction that the qubits involved are not error-protected. Consequently, idling errors must be considered. Since pipeline performance depends on the degree of parallelisation, idling depends on available memory, slightly complicating the analysis. However, by full numerical analysis we find a simple outcome: for distilling from 5\% to 1\%, the optimal strategy across nearly the entire parameter space is two rounds of the classical $[2,1,2]$ parity code.

\subsection{Networking Regimes Across Platforms}

The primary difference between the two plots in \cref{fig:top view} lies in the performance of the direct-gate methods: both degrade more rapidly in the lower-quality case ($p_{\mathrm{bell}} = 5\%$), making logical distillation far more dominant. This occurs because direct methods are highly sensitive to initial error rates; reaching target fidelities under poor conditions requires much larger logical codes, reducing rates significantly. Idling further amplifies these effects.

In general, transversal gates tend to outperform other methods when the physical Bell pair rate $r_{\mathrm{bell}}$ is high. When $r_{\mathrm{bell}}$ is low and system sizes are small, lattice surgery is typically the only viable approach. For large systems operating at moderate to low values of $r_{\mathrm{bell}}$, logical distillation emerges as the most effective method. In the lower-quality case, around when $r_{\mathrm{bell}}/r_{\mathrm{physical}} < 0.2$, physical distillation to 1\% becomes infeasible. As a result, neither direct-gate method can reach the target error rate, further emphasising the necessity of logical distillation across a large portion of parameter space.

Regarding the platforms in \cref{sec:platforms}, we emphasise that the figure does not compare absolute distributed gate rates, as all quantities are normalised to local rates. Instead, it illustrates the \emph{relative impact} of modular connections. We collect our observations as such:

\paragraph{Neutral atoms.} The region spans all three regimes in both plots. This reflects relatively slow local gates and high potential for Bell pair multiplexing, leading to high $r_{\mathrm{bell}}/r_{\mathrm{physical}}$. Modular approaches can match or exceed local performance, making this platform  suited for distributed architectures.

\paragraph{Superconducting qubits.} Fast local gates shift the region to the left, where only distillation is viable due to severe idling penalties. This suggests that while limited modularity is acceptable, particularly when only a few distributed gates are needed, scaling up the number of modular links quickly turns into a bottleneck. In such cases, the slower distributed operations lag behind fast local gates, negatively impacting performance.

\paragraph{SiV cavities.} This hardware platform is characterised by the similar parameters for local and non-local gates.  For high-quality Bell pairs, lattice surgery emerges as the only feasible method while for lower-quality input, logical distillation dominates. Since the physical Bell pair rate is bounded by the local gate rate, transversal gates are not suitable for this hardware.

% \begin{itemize}
%     \item \textbf{Neutral atoms}: Their region spans all three regimes in both plots. This reflects relatively slow local gates and high potential for Bell pair multiplexing, leading to high $r_{\mathrm{bell}}/r_{\mathrm{physical}}$. Modular approaches can match or exceed local performance, making this platform especially suited for distributed architectures.

%     \item \textbf{Superconducting qubits}: Fast local gates shift the region to the left, where only distillation is viable due to severe idling penalties. This suggests that while limited modularity is acceptable, particularly when only a few distributed gates are needed, scaling up the number of modular links quickly turns into a bottleneck. In such cases, the slower distributed operations lag behind fast local gates, negatively impacting performance.

%     \item \textbf{SiV cavities}: The regime differs significantly between plots. For high-quality Bell pairs, only lattice surgery is feasible. For lower-quality input, distillation dominates. The significant space requirements suggest that current methods provide limited advantage for SiV, and alternative networking strategies should be explored.
% \end{itemize}

\section{Conclusion}
In summary, we have introduced a novel grow-and-distil sequence reducing qubit overhead by several thousand while enhancing logical Bell pair generation rates compared to previous approaches to logical Bell pair distillation. Additionally, we have developed a systematic framework for cross-method comparison and optimisation of fault-tolerant interfaces between QPUs including lattice surgery and transversal gate techniques. Focusing on the surface code, we mapped the regimes of qubit coherence times, local gate fidelities and speeds, network entanglement rates and fidelities, and intra-QPU qubit resources where grow-and-distil, lattice surgery, and transversal gates maximise logical Bell pair generation. By adopting future projected parameter regimes for superconducting, atomic, and solid-state platforms, we identified both the required number of networking qubits and the optimal interface strategies to reach target logical error rates of $10^{-6}$ and $10^{-12}$.

These results provide a reference for experimental prioritisation and architectural design in scalable modular quantum computing. Specifically, for the surface code, matching inter-module operation rates to intra-module gates will typically require several thousand networking qubits and physical entangling rates up to three orders of magnitude faster than local gate speeds. These stringent requirements are largely dictated by the encoding rate of the QEC code, suggesting that adopting more efficient codes such as qLDPC codes \cite{ataides2025,yoder2025}, which maintain high error thresholds while improving the logical encoding ratio could significantly lower the requirements for modular computation. A higher logical encoding rate would reduce the qubit overhead of the grow-and-distil techniques and similarly decrease the entanglement distribution rates required for transversal gates, since fewer physical qubits are needed per logical qubit. Consequently, while we expect the overall trend of transversal gates being optimal for high entangling rates, lattice surgery for intermediate rates, and grow-and-distill for slower rates to persist, as shown in \cref{fig:top view}, the axis would shift towards lower memory and physical entangling rates. 

Our analysis is intentionally platform-agnostic, focusing on universal design principles that offer a reference for future hardware developments. A natural next step would be to incorporate more specific hardware constraints into the optimization. For instance, in our current simulations, we assume fixed rates and error parameters for local gates and syndrome extraction as the size of the qubit memories increased. In practice, however, larger qubit memories may require longer intra-QPU transport times and could introduce additional errors affecting the performance. While beyond the scope of this work, exploring these effects in detail would be an interesting direction for future study. 

\section{Acknowledgements}
We would like to thank Mikhail Lukin, Josiah Sinclair, Madelyn Cain, and David Levonian for helpful and stimulating discussions and general support. We gratefully acknowledge support from Innovation Fund Denmark under grant no. 1063-00046B - “PhotoQ Photonic Quantum Computing” and The AWS Quantum Discovery Fund at the Harvard Quantum Initiative. G.B. acknowledges support from the MIT Patrons of Physics Fellows Society.

\section{Code availability}
The code used for producing our numerical results as well as that used for analysing and plotting of the data is available at ref. \cite{marqversen_quantum_computations_2025}, and the code used for validating our model of surface code growing and Bell pair injection at ref. \cite{sirotin_growing_injection_2025}.

\printbibliography

@misc{yoder2025,
      title={Tour de gross: A modular quantum computer based on bivariate bicycle codes}, 
      author={Theodore J. Yoder and Eddie Schoute and Patrick Rall and Emily Pritchett and Jay M. Gambetta and Andrew W. Cross and Malcolm Carroll and Michael E. Beverland},
      year={2025},
      eprint={2506.03094},
      archivePrefix={arXiv},
      primaryClass={quant-ph},
      url={https://arxiv.org/abs/2506.03094}, 
}

@article{bluvstein2025architectural,
  title={Architectural mechanisms of a universal fault-tolerant quantum computer},
  author={Bluvstein, Dolev and Geim, Alexandra A and Li, Sophie H and Evered, Simon J and Ataides, J and Baranes, Gefen and Gu, Andi and Manovitz, Tom and Xu, Muqing and Kalinowski, Marcin and others},
  journal={arXiv preprint arXiv:2506.20661},
  year={2025}
}

@article{Monroe2014,
  title = {Large-scale modular quantum-computer architecture with atomic memory and photonic interconnects},
  author = {Monroe, C. and Raussendorf, R. and Ruthven, A. and Brown, K. R. and Maunz, P. and Duan, L.-M. and Kim, J.},
  journal = {Phys. Rev. A},
  volume = {89},
  issue = {2},
  pages = {022317},
  numpages = {16},
  year = {2014},
  month = {2},
  publisher = {American Physical Society},
  doi = {10.1103/PhysRevA.89.022317},
  url = {https://link.aps.org/doi/10.1103/PhysRevA.89.022317}
}

@misc{stack2025,
      title={Assessing Teleportation of Logical Qubits in a Distributed Quantum Architecture under Error Correction}, 
      author={John Stack and Ming Wang and Frank Mueller},
      year={2025},
      eprint={2504.05611},
      archivePrefix={arXiv},
      primaryClass={quant-ph},
      url={https://arxiv.org/abs/2504.05611}, 
}

@inproceedings{pattison_fast_2024,
author = {Pattison, Christopher and Baranes, Gefen and Bonilla Ataides, Juan Pablo and Lukin, Mikhail D. and Zhou, Hengyun},
title = {Constant-Rate Entanglement Distillation for Fast Quantum Interconnects},
year = {2025},
isbn = {9798400712616},
publisher = {Association for Computing Machinery},
address = {New York, NY, USA},
url = {https://doi.org/10.1145/3695053.3731069},
doi = {10.1145/3695053.3731069},
booktitle = {Proceedings of the 52nd Annual International Symposium on Computer Architecture},
pages = {257–270},
numpages = {14},
location = {
},
series = {ISCA '25}
}

@article{ramette_fault-tolerant_2024,
    title = {Fault-tolerant connection of error-corrected qubits with noisy links},
    volume = {10},
    copyright = {2024 The Author(s)},
    issn = {2056-6387},
    url = {https://www.nature.com/articles/s41534-024-00855-4},
    doi = {10.1038/s41534-024-00855-4},
    language = {en},
    number = {1},
    urldate = {2025-03-12},
    journal = {npj Quantum Information},
    author = {Ramette, Joshua and Sinclair, Josiah and Breuckmann, Nikolas P. and Vuletić, Vladan},
    month = jun,
    year = {2024},
    note = {Publisher: Nature Publishing Group},
    pages = {1--6},
}

@inproceedings{lao_magic_2022,
    address = {Turin Italy},
    title = {Magic state injection on the rotated surface code},
    isbn = {978-1-4503-9338-6},
    url = {https://dl.acm.org/doi/10.1145/3528416.3530237},
    doi = {10.1145/3528416.3530237},
    language = {en},
    urldate = {2025-03-25},
    booktitle = {Proceedings of the 19th {ACM} {International} {Conference} on {Computing} {Frontiers}},
    publisher = {ACM},
    author = {Lao, Lingling and Criger, Ben},
    month = may,
    year = {2022},
    pages = {113--120},
}

@misc{cain_fast_2025,
    title = {Fast correlated decoding of transversal logical algorithms},
    url = {http://arxiv.org/abs/2505.13587},
    doi = {10.48550/arXiv.2505.13587},
    urldate = {2025-05-22},
    publisher = {arXiv},
    author = {Cain, Madelyn and Bluvstein, Dolev and Zhao, Chen and Gu, Shouzhen and Maskara, Nishad and Kalinowski, Marcin and Geim, Alexandra A. and Kubica, Aleksander and Lukin, Mikhail D. and Zhou, Hengyun},
    month = may,
    year = {2025},
    note = {arXiv:2505.13587 [quant-ph]},
}

@article{cain_correlated_2024,
    title = {Correlated {Decoding} of {Logical} {Algorithms} with {Transversal} {Gates}},
    volume = {133},
    url = {https://link.aps.org/doi/10.1103/PhysRevLett.133.240602},
    doi = {10.1103/PhysRevLett.133.240602},
    number = {24},
    urldate = {2025-06-19},
    journal = {Physical Review Letters},
    author = {Cain, Madelyn and Zhao, Chen and Zhou, Hengyun and Meister, Nadine and Ataides, J. Pablo Bonilla and Jaffe, Arthur and Bluvstein, Dolev and Lukin, Mikhail D.},
    month = dec,
    year = {2024},
    note = {Publisher: American Physical Society},
    pages = {240602},
}

@misc{zhou_algorithmic_2024,
    title = {Algorithmic {Fault} {Tolerance} for {Fast} {Quantum} {Computing}},
    url = {https://arxiv.org/abs/2406.17653},
    language = {en},
    publisher = {arXiv},
    author = {Zhou, Hengyun and Zhao, Chen and Cain, Madelyn and Bluvstein, Dolev and Duckering, Casey and Hu, Hong-Ye and Wang, Sheng-Tao and Kubica, Aleksander and Lukin, Mikhail D},
    month = jun,
    year = {2024},
    note = {arXiv:2406.17653 [quant-ph]},
}

@article{li_magic_2015,
    title = {A magic state’s fidelity can be superior to the operations that created it},
    volume = {17},
    issn = {1367-2630},
    url = {https://iopscience.iop.org/article/10.1088/1367-2630/17/2/023037},
    doi = {10.1088/1367-2630/17/2/023037},
    language = {en},
    number = {2},
    urldate = {2025-04-24},
    journal = {New Journal of Physics},
    author = {Li, Ying},
    month = feb,
    year = {2015},
    pages = {023037},
}

@article{baranes2025designing,
  title={Designing Fault-Tolerant Blind Quantum Computation},
  author={Baranes, Gefen and Wang, Iria W and Machado, Francisco and Suleymanzade, Aziza and Stas, Pieter-Jan and Wei, Yan-Cheng and Yelin, Susanne F and Borregaard, Johannes and Lukin, Mikhail D},
  journal={arXiv preprint arXiv:2505.21621},
  year={2025}
}

@article{sinclair_2025neutral_atoms_interconnect,
  title={Fault-tolerant optical interconnects for neutral-atom arrays},
  author={Sinclair, Josiah and Ramette, Joshua and Grinkemeyer, Brandon and Bluvstein, Dolev and Lukin, Mikhail D and Vuleti{\'c}, Vladan},
  journal={Physical Review Research},
  volume={7},
  number={1},
  pages={013313},
  year={2025},
  publisher={APS}
}

@article{Li_2025neutral_atoms_interconnect,
  title = {High-Rate and High-Fidelity Modular Interconnects between Neutral Atom Quantum Processors},
  author = {Li, Yiyi and Thompson, Jeff D.},
  journal = {PRX Quantum},
  volume = {5},
  issue = {2},
  pages = {020363},
  numpages = {13},
  year = {2024},
  month = {6},
  publisher = {American Physical Society},
  doi = {10.1103/PRXQuantum.5.020363},
  url = {https://link.aps.org/doi/10.1103/PRXQuantum.5.020363}
}

@article{knaut_entanglement_2024,
    title = {Entanglement of nanophotonic quantum memory nodes in a telecom network},
    volume = {629},
    issn = {1476-4687},
    url = {https://doi.org/10.1038/s41586-024-07252-z},
    doi = {10.1038/s41586-024-07252-z},
    number = {8012},
    journal = {Nature},
    author = {Knaut, C. M. and Suleymanzade, A. and Wei, Y.-C. and Assumpcao, D. R. and Stas, P.-J. and Huan, Y. Q. and Machielse, B. and Knall, E. N. and Sutula, M. and Baranes, G. and Sinclair, N. and De-Eknamkul, C. and Levonian, D. S. and Bhaskar, M. K. and Park, H. and Lončar, M. and Lukin, M. D.},
    month = may,
    year = {2024},
    pages = {573--578},
}

@article{manetsch2024tweezer,
  title={A tweezer array with 6100 highly coherent atomic qubits},
  author={Manetsch, Hannah J and Nomura, Gyohei and Bataille, Elie and Leung, Kon H and Lv, Xudong and Endres, Manuel},
  journal={arXiv preprint arXiv:2403.12021},
  year={2024}
}

@article{Bluvstein_2024_neutralatomQC,
	author = {Bluvstein, Dolev and Evered, Simon J. and Geim, Alexandra A. and Li, Sophie H. and Zhou, Hengyun and Manovitz, Tom and Ebadi, Sepehr and Cain, Madelyn and Kalinowski, Marcin and Hangleiter, Dominik and Bonilla Ataides, J. Pablo and Maskara, Nishad and Cong, Iris and Gao, Xun and Sales Rodriguez, Pedro and Karolyshyn, Thomas and Semeghini, Giulia and Gullans, Michael J. and Greiner, Markus and Vuleti{\'c}, Vladan and Lukin, Mikhail D.},
	doi = {10.1038/s41586-023-06927-3},
	id = {Bluvstein2024},
	journal = {Nature},
	number = {7997},
	pages = {58--65},
	title = {Logical quantum processor based on reconfigurable atom arrays},
	url = {https://doi.org/10.1038/s41586-023-06927-3},
	volume = {626},
	year = {2024}}

@article{google_2024_below_threshold_SC,
	author = {Acharya, Rajeev and Abanin, Dmitry A. and Aghababaie-Beni, Laleh and Aleiner, Igor and Andersen, Trond I. and Ansmann, Markus and Arute, Frank and Arya, Kunal and Asfaw, Abraham and Astrakhantsev, Nikita and Atalaya, Juan and Babbush, Ryan and Bacon, Dave and Ballard, Brian and Bardin, Joseph C. and Bausch, Johannes and Bengtsson, Andreas and Bilmes, Alexander and Blackwell, Sam and Boixo, Sergio and Bortoli, Gina and Bourassa, Alexandre and Bovaird, Jenna and Brill, Leon and Broughton, Michael and Browne, David A. and Buchea, Brett and Buckley, Bob B. and Buell, David A. and Burger, Tim and Burkett, Brian and Bushnell, Nicholas and Cabrera, Anthony and Campero, Juan and Chang, Hung-Shen and Chen, Yu and Chen, Zijun and Chiaro, Ben and Chik, Desmond and Chou, Charina and Claes, Jahan and Cleland, Agnetta Y. and Cogan, Josh and Collins, Roberto and Conner, Paul and Courtney, William and Crook, Alexander L. and Curtin, Ben and Das, Sayan and Davies, Alex and De Lorenzo, Laura and Debroy, Dripto M. and Demura, Sean and Devoret, Michel and Di Paolo, Agustin and Donohoe, Paul and Drozdov, Ilya and Dunsworth, Andrew and Earle, Clint and Edlich, Thomas and Eickbusch, Alec and Elbag, Aviv Moshe and Elzouka, Mahmoud and Erickson, Catherine and Faoro, Lara and Farhi, Edward and Ferreira, Vinicius S. and Burgos, Leslie Flores and Forati, Ebrahim and Fowler, Austin G. and Foxen, Brooks and Ganjam, Suhas and Garcia, Gonzalo and Gasca, Robert and Genois, {\'E}lie and Giang, William and Gidney, Craig and Gilboa, Dar and Gosula, Raja and Dau, Alejandro Grajales and Graumann, Dietrich and Greene, Alex and Gross, Jonathan A. and Habegger, Steve and Hall, John and Hamilton, Michael C. and Hansen, Monica and Harrigan, Matthew P. and Harrington, Sean D. and Heras, Francisco J. H. and Heslin, Stephen and Heu, Paula and Higgott, Oscar and Hill, Gordon and Hilton, Jeremy and Holland, George and Hong, Sabrina and Huang, Hsin-Yuan and Huff, Ashley and Huggins, William J. and Ioffe, Lev B. and Isakov, Sergei V. and Iveland, Justin and Jeffrey, Evan and Jiang, Zhang and Jones, Cody and Jordan, Stephen and Joshi, Chaitali and Juhas, Pavol and Kafri, Dvir and Kang, Hui and Karamlou, Amir H. and Kechedzhi, Kostyantyn and Kelly, Julian and Khaire, Trupti and Khattar, Tanuj and Khezri, Mostafa and Kim, Seon and Klimov, Paul V. and Klots, Andrey R. and Kobrin, Bryce and Kohli, Pushmeet and Korotkov, Alexander N. and Kostritsa, Fedor and Kothari, Robin and Kozlovskii, Borislav and Kreikebaum, John Mark and Kurilovich, Vladislav D. and Lacroix, Nathan and Landhuis, David and Lange-Dei, Tiano and Langley, Brandon W. and Laptev, Pavel and Lau, Kim-Ming and Le Guevel, Lo{\"\i}ck and Ledford, Justin and Lee, Joonho and Lee, Kenny and Lensky, Yuri D. and Leon, Shannon and Lester, Brian J. and Li, Wing Yan and Li, Yin and Lill, Alexander T. and Liu, Wayne and Livingston, William P. and Locharla, Aditya and Lucero, Erik and Lundahl, Daniel and Lunt, Aaron and Madhuk, Sid and Malone, Fionn D. and Maloney, Ashley and Mandr{\`a}, Salvatore and Manyika, James and Martin, Leigh S. and Martin, Orion and Martin, Steven and Maxfield, Cameron and McClean, Jarrod R. and McEwen, Matt and Meeks, Seneca and Megrant, Anthony and Mi, Xiao and Miao, Kevin C. and Mieszala, Amanda and Molavi, Reza and Molina, Sebastian and Montazeri, Shirin and Morvan, Alexis and Movassagh, Ramis and Mruczkiewicz, Wojciech and Naaman, Ofer and Neeley, Matthew and Neill, Charles and Nersisyan, Ani and Neven, Hartmut and Newman, Michael and Ng, Jiun How and Nguyen, Anthony and Nguyen, Murray and Ni, Chia-Hung and Niu, Murphy Yuezhen and O'Brien, Thomas E. and Oliver, William D. and Opremcak, Alex and Ottosson, Kristoffer and Petukhov, Andre and Pizzuto, Alex and Platt, John and Potter, Rebecca and Pritchard, Orion and Pryadko, Leonid P. and Quintana, Chris and Ramachandran, Ganesh and Reagor, Matthew J. and Redding, John and Rhodes, David M. and Roberts, Gabrielle and Rosenberg, Eliott and Rosenfeld, Emma and Roushan, Pedram and Rubin, Nicholas C. and Saei, Negar and Sank, Daniel and Sankaragomathi, Kannan and Satzinger, Kevin J. and Schurkus, Henry F. and Schuster, Christopher and Senior, Andrew W. and Shearn, Michael J. and Shorter, Aaron and Shutty, Noah and Shvarts, Vladimir and Singh, Shraddha and Sivak, Volodymyr and Skruzny, Jindra and Small, Spencer and Smelyanskiy, Vadim and Smith, W. Clarke and Somma, Rolando D. and Springer, Sofia and Sterling, George and Strain, Doug and Suchard, Jordan and Szasz, Aaron and Sztein, Alex and Thor, Douglas and Torres, Alfredo and Torunbalci, M. Mert and Vaishnav, Abeer and Vargas, Justin and Vdovichev, Sergey and Vidal, Guifre and Villalonga, Benjamin and Heidweiller, Catherine Vollgraff and Waltman, Steven and Wang, Shannon X. and Ware, Brayden and Weber, Kate and Weidel, Travis and White, Theodore and Wong, Kristi and Woo, Bryan W. K. and Xing, Cheng and Yao, Z. Jamie and Yeh, Ping and Ying, Bicheng and Yoo, Juhwan and Yosri, Noureldin and Young, Grayson and Zalcman, Adam and Zhang, Yaxing and Zhu, Ningfeng and Zobrist, Nicholas and Google Quantum AI and Collaborators},
	doi = {10.1038/s41586-024-08449-y},
	id = {Acharya2025},
	journal = {Nature},
	number = {8052},
	pages = {920--926},
	title = {Quantum error correction below the surface code threshold},
	url = {https://doi.org/10.1038/s41586-024-08449-y},
	volume = {638},
	year = {2025}}

@article{Covey_neutral_atoms_quantum_networks,
	author = {Covey, Jacob P. and Weinfurter, Harald and Bernien, Hannes},
	doi = {10.1038/s41534-023-00759-9},
	id = {Covey2023},
	journal = {npj Quantum Information},
	number = {1},
	pages = {90},
	title = {Quantum networks with neutral atom processing nodes},
	url = {https://doi.org/10.1038/s41534-023-00759-9},
	volume = {9},
	year = {2023}}

@article{Evered_high_fidelity_gate_atoms,
	author = {Evered, Simon J. and Bluvstein, Dolev and Kalinowski, Marcin and Ebadi, Sepehr and Manovitz, Tom and Zhou, Hengyun and Li, Sophie H. and Geim, Alexandra A. and Wang, Tout T. and Maskara, Nishad and Levine, Harry and Semeghini, Giulia and Greiner, Markus and Vuleti{\'c}, Vladan and Lukin, Mikhail D.},
	doi = {10.1038/s41586-023-06481-y},
	id = {Evered2023},
	journal = {Nature},
	number = {7982},
	pages = {268--272},
	title = {High-fidelity parallel entangling gates on a neutral-atom quantum computer},
	url = {https://doi.org/10.1038/s41586-023-06481-y},
	volume = {622},
	year = {2023}}

@article{Cheng_SiV_BQC,
	author = {Y.-C. Wei and P.-J. Stas and A. Suleymanzade and G. Baranes and F. Machado and Y. Q. Huan and C. M. Knaut and S. W. Ding and M. Merz and E. N. Knall and U. Yazlar and M. Sirotin and I. W. Wang and B. Machielse and S. F. Yelin and J. Borregaard and H. Park and M. Lon{\v c}ar and M. D. Lukin},
	doi = {10.1126/science.adu6894},
	journal = {Science},
	number = {6746},
	pages = {509-513},
	title = {Universal distributed blind quantum computing with solid-state qubits},
	url = {https://www.science.org/doi/abs/10.1126/science.adu6894},
	volume = {388},
	year = {2025}}

@article{acharya_suppressing_2023,
    title = {Suppressing quantum errors by scaling a surface code logical qubit},
    volume = {614},
    issn = {1476-4687},
    url = {https://doi.org/10.1038/s41586-022-05434-1},
    doi = {10.1038/s41586-022-05434-1},
    number = {7949},
    journal = {Nature},
    author = {Acharya, Rajeev and Aleiner, Igor and Allen, Richard and Andersen, Trond I. and Ansmann, Markus and Arute, Frank and Arya, Kunal and Asfaw, Abraham and Atalaya, Juan and Babbush, Ryan and Bacon, Dave and Bardin, Joseph C. and Basso, Joao and Bengtsson, Andreas and Boixo, Sergio and Bortoli, Gina and Bourassa, Alexandre and Bovaird, Jenna and Brill, Leon and Broughton, Michael and Buckley, Bob B. and Buell, David A. and Burger, Tim and Burkett, Brian and Bushnell, Nicholas and Chen, Yu and Chen, Zijun and Chiaro, Ben and Cogan, Josh and Collins, Roberto and Conner, Paul and Courtney, William and Crook, Alexander L. and Curtin, Ben and Debroy, Dripto M. and Del Toro Barba, Alexander and Demura, Sean and Dunsworth, Andrew and Eppens, Daniel and Erickson, Catherine and Faoro, Lara and Farhi, Edward and Fatemi, Reza and Flores Burgos, Leslie and Forati, Ebrahim and Fowler, Austin G. and Foxen, Brooks and Giang, William and Gidney, Craig and Gilboa, Dar and Giustina, Marissa and Grajales Dau, Alejandro and Gross, Jonathan A. and Habegger, Steve and Hamilton, Michael C. and Harrigan, Matthew P. and Harrington, Sean D. and Higgott, Oscar and Hilton, Jeremy and Hoffmann, Markus and Hong, Sabrina and Huang, Trent and Huff, Ashley and Huggins, William J. and Ioffe, Lev B. and Isakov, Sergei V. and Iveland, Justin and Jeffrey, Evan and Jiang, Zhang and Jones, Cody and Juhas, Pavol and Kafri, Dvir and Kechedzhi, Kostyantyn and Kelly, Julian and Khattar, Tanuj and Khezri, Mostafa and Kieferová, Mária and Kim, Seon and Kitaev, Alexei and Klimov, Paul V. and Klots, Andrey R. and Korotkov, Alexander N. and Kostritsa, Fedor and Kreikebaum, John Mark and Landhuis, David and Laptev, Pavel and Lau, Kim-Ming and Laws, Lily and Lee, Joonho and Lee, Kenny and Lester, Brian J. and Lill, Alexander and Liu, Wayne and Locharla, Aditya and Lucero, Erik and Malone, Fionn D. and Marshall, Jeffrey and Martin, Orion and McClean, Jarrod R. and McCourt, Trevor and McEwen, Matt and Megrant, Anthony and Meurer Costa, Bernardo and Mi, Xiao and Miao, Kevin C. and Mohseni, Masoud and Montazeri, Shirin and Morvan, Alexis and Mount, Emily and Mruczkiewicz, Wojciech and Naaman, Ofer and Neeley, Matthew and Neill, Charles and Nersisyan, Ani and Neven, Hartmut and Newman, Michael and Ng, Jiun How and Nguyen, Anthony and Nguyen, Murray and Niu, Murphy Yuezhen and O’Brien, Thomas E. and Opremcak, Alex and Platt, John and Petukhov, Andre and Potter, Rebecca and Pryadko, Leonid P. and Quintana, Chris and Roushan, Pedram and Rubin, Nicholas C. and Saei, Negar and Sank, Daniel and Sankaragomathi, Kannan and Satzinger, Kevin J. and Schurkus, Henry F. and Schuster, Christopher and Shearn, Michael J. and Shorter, Aaron and Shvarts, Vladimir and Skruzny, Jindra and Smelyanskiy, Vadim and Smith, W. Clarke and Sterling, George and Strain, Doug and Szalay, Marco and Torres, Alfredo and Vidal, Guifre and Villalonga, Benjamin and Vollgraff Heidweiller, Catherine and White, Theodore and Xing, Cheng and Yao, Z. Jamie and Yeh, Ping and Yoo, Juhwan and Young, Grayson and Zalcman, Adam and Zhang, Yaxing and Zhu, Ningfeng and {Google Quantum AI}},
    month = feb,
    year = {2023},
    pages = {676--681},
}

@article{Wallraff_SC_entanglement_BellTest,
	author = {Storz, Simon and Sch{\"a}r, Josua and Kulikov, Anatoly and Magnard, Paul and Kurpiers, Philipp and L{\"u}tolf, Janis and Walter, Theo and Copetudo, Adrian and Reuer, Kevin and Akin, Abdulkadir and Besse, Jean-Claude and Gabureac, Mihai and Norris, Graham J. and Rosario, Andr{\'e}s and Martin, Ferran and Martinez, Jos{\'e} and Amaya, Waldimar and Mitchell, Morgan W. and Abellan, Carlos and Bancal, Jean-Daniel and Sangouard, Nicolas and Royer, Baptiste and Blais, Alexandre and Wallraff, Andreas},
	doi = {10.1038/s41586-023-05885-0},
	id = {Storz2023},
	journal = {Nature},
	number = {7960},
	pages = {265--270},
	title = {Loophole-free Bell inequality violation with superconducting circuits},
	url = {https://doi.org/10.1038/s41586-023-05885-0},
	volume = {617},
	year = {2023}}

@article{bartling2024universal,
title = {Universal high-fidelity quantum gates for spin qubits in diamond},
author = {Bartling, H.P. and Yun, J. and Schymik, K.N. and van Riggelen, M. and Enthoven, L.A. and van Ommen, H.B. and Babaie, M. and Sebastiano, F. and Markham, M. and Twitchen, D.J. and Taminiau, T.H.},
journal = {Phys. Rev. Appl.},
volume = {23},
issue = {3},
pages = {034052},
numpages = {26},
year = {2025},
month = {Mar},
publisher = {American Physical Society},
doi = {10.1103/PhysRevApplied.23.034052},
url = {https://link.aps.org/doi/10.1103/PhysRevApplied.23.034052}
}

@article{stas2022robust,
  title={Robust multi-qubit quantum network node with integrated error detection},
  author={Stas, P-J and Huan, Yan Qi and Machielse, Bartholomeus and Knall, Erik N and Suleymanzade, Aziza and Pingault, Benjamin and Sutula, Madison and Ding, Sophie W and Knaut, Can M and Assumpcao, Daniel R and others},
  journal={Science},
  volume={378},
  number={6619},
  pages={557--560},
  year={2022},
  publisher={American Association for the Advancement of Science}
}

@book{majidy2024building,
  title={Building Quantum Computers: A Practical Introduction},
  author={Majidy, Shayan and Wilson, Christopher and Laflamme, Raymond},
  year={2024},
  publisher={Cambridge University Press}
}

@article{chiu2025continuous,
  title={Continuous operation of a coherent 3,000-qubit system},
  author={Chiu, Neng-Chun and Trapp, Elias C and Guo, Jinen and Abobeih, Mohamed H and Stewart, Luke M and Hollerith, Simon and Stroganov, Pavel and Kalinowski, Marcin and Geim, Alexandra A and Evered, Simon J and others},
  journal={arXiv preprint arXiv:2506.20660},
  year={2025}
}

@article{ritter2012elementary,
  title={An elementary quantum network of single atoms in optical cavities},
  author={Ritter, Stephan and N{\"o}lleke, Christian and Hahn, Carolin and Reiserer, Andreas and Neuzner, Andreas and Uphoff, Manuel and M{\"u}cke, Martin and Figueroa, Eden and Bochmann, Joerg and Rempe, Gerhard},
  journal={Nature},
  volume={484},
  number={7393},
  pages={195--200},
  year={2012},
  publisher={Nature Publishing Group UK London}
}

@article{fowler2012surface,
  title={Surface codes: Towards practical large-scale quantum computation},
  author={Fowler, Austin G and Mariantoni, Matteo and Martinis, John M and Cleland, Andrew N},
  journal={Physical Review A—Atomic, Molecular, and Optical Physics},
  volume={86},
  number={3},
  pages={032324},
  year={2012},
  publisher={APS}
}

@article{baranes2025leveraging,
  title={Leveraging atom loss errors in fault tolerant quantum algorithms},
  author={Baranes, Gefen and Cain, Madelyn and Ataides, J and Bluvstein, Dolev and Sinclair, Josiah and Vuletic, Vladan and Zhou, Hengyun and Lukin, Mikhail D},
  journal={arXiv preprint arXiv:2502.20558},
  year={2025}
}

@article{Krastanov_PRL_2021_SC_modular,
  title = {Optically Heralded Entanglement of Superconducting Systems in Quantum Networks},
  author = {Krastanov, Stefan and Raniwala, Hamza and Holzgrafe, Jeffrey and Jacobs, Kurt and Lon\ifmmode \check{c}\else \v{c}\fi{}ar, Marko and Reagor, Matthew J. and Englund, Dirk R.},
  journal = {Phys. Rev. Lett.},
  volume = {127},
  issue = {4},
  pages = {040503},
  numpages = {7},
  year = {2021},
  month = {7},
  publisher = {American Physical Society},
  doi = {10.1103/PhysRevLett.127.040503},
  url = {https://link.aps.org/doi/10.1103/PhysRevLett.127.040503}
}

@misc{marqversen_quantum_computations_2025,
    title = {quantum\_computations},
    copyright = {MIT},
    url = {https://github.com/frederik-kofoed-marqversen/quantum_computations},
    urldate = {2025-03-03},
    author = {Marqversen, Frederik Kofoed},
    month = mar,
    year = {2025},
    note = {original-date: 2024-08-29},
    howpublished = {\url{https://github.com/frederik-kofoed-marqversen/quantum_computations}},
}

@misc{sirotin_growing_injection_2025,
    title = {Growing\_simulation},
    copyright = {MIT},
    url = {https://github.com/MaximSirotin/BellPairsCode},
    urldate = {2025-07-07},
    author = {Sirotin, Maxim},
    month = jul,
    year = {2025},
    note = {original-date: 2025-07-07},
    howpublished = {\url{https://github.com/MaximSirotin/BellPairsCode}},
}

@article{gidney2021stim,
  title={Stim: a fast stabilizer circuit simulator},
  author={Gidney, Craig},
  journal={Quantum},
  volume={5},
  pages={497},
  year={2021},
  publisher={Verein zur F{\"o}rderung des Open Access Publizierens in den Quantenwissenschaften}
}

@article{Lee2021,
  title = {Even More Efficient Quantum Computations of Chemistry Through Tensor Hypercontraction},
  author = {Lee, Joonho and Berry, Dominic W. and Gidney, Craig and Huggins, William J. and McClean, Jarrod R. and Wiebe, Nathan and Babbush, Ryan},
  journal = {PRX Quantum},
  volume = {2},
  issue = {3},
  pages = {030305},
  numpages = {62},
  year = {2021},
  month = {7},
  publisher = {American Physical Society},
  doi = {10.1103/PRXQuantum.2.030305},
  url = {https://link.aps.org/doi/10.1103/PRXQuantum.2.030305}
}

@misc{ataides2025,
      title={Constant-Overhead Fault-Tolerant Bell-Pair Distillation using High-Rate Codes}, 
      author={J. Pablo Bonilla Ataides and Hengyun Zhou and Qian Xu and Gefen Baranes and Bikun Li and Mikhail D. Lukin and Liang Jiang},
      year={2025},
      eprint={2502.09542},
      archivePrefix={arXiv},
      primaryClass={quant-ph},
      url={https://arxiv.org/abs/2502.09542}, 
}

@misc{beverland2022,
      title={Assessing requirements to scale to practical quantum advantage}, 
      author={Michael E. Beverland and Prakash Murali and Matthias Troyer and Krysta M. Svore and Torsten Hoefler and Vadym Kliuchnikov and Guang Hao Low and Mathias Soeken and Aarthi Sundaram and Alexander Vaschillo},
      year={2022},
      eprint={2211.07629},
      archivePrefix={arXiv},
      primaryClass={quant-ph},
      url={https://arxiv.org/abs/2211.07629}, 
}

@misc{dalzell2023,
      title={Quantum algorithms: A survey of applications and end-to-end complexities}, 
      author={Alexander M. Dalzell and Sam McArdle and Mario Berta and Przemyslaw Bienias and Chi-Fang Chen and András Gilyén and Connor T. Hann and Michael J. Kastoryano and Emil T. Khabiboulline and Aleksander Kubica and Grant Salton and Samson Wang and Fernando G. S. L. Brandão},
      year={2023},
      eprint={2310.03011},
      archivePrefix={arXiv},
      primaryClass={quant-ph},
      url={https://arxiv.org/abs/2310.03011}, 
}

@article{Gupta2024,
	author = {Gupta, Riddhi S. and Sundaresan, Neereja and Alexander, Thomas and Wood, Christopher J. and Merkel, Seth T. and Healy, Michael B. and Hillenbrand, Marius and Jochym-O'Connor, Tomas and Wootton, James R. and Yoder, Theodore J. and Cross, Andrew W. and Takita, Maika and Brown, Benjamin J.},
	date = {2024-01-01},
	doi = {10.1038/s41586-023-06846-3},
	id = {Gupta2024},
	journal = {Nature},
	number = {7994},
	pages = {259--263},
	title = {Encoding a magic state with beyond break-even fidelity},
	url = {https://doi.org/10.1038/s41586-023-06846-3},
	volume = {625},
	year = {2024}}

@article{Putterman2025,
	author = {Putterman, Harald and Noh, Kyungjoo and Hann, Connor T. and MacCabe, Gregory S. and Aghaeimeibodi, Shahriar and Patel, Rishi N. and Lee, Menyoung and Jones, William M. and Moradinejad, Hesam and Rodriguez, Roberto and Mahuli, Neha and Rose, Jefferson and Owens, John Clai and Levine, Harry and Rosenfeld, Emma and Reinhold, Philip and Moncelsi, Lorenzo and Alcid, Joshua Ari and Alidoust, Nasser and Arrangoiz-Arriola, Patricio and Barnett, James and Bienias, Przemyslaw and Carson, Hugh A. and Chen, Cliff and Chen, Li and Chinkezian, Harutiun and Chisholm, Eric M. and Chou, Ming-Han and Clerk, Aashish and Clifford, Andrew and Cosmic, R. and Curiel, Ana Valdes and Davis, Erik and DeLorenzo, Laura and D'Ewart, J. Mitchell and Diky, Art and D'Souza, Nathan and Dumitrescu, Philipp T. and Eisenmann, Shmuel and Elkhouly, Essam and Evenbly, Glen and Fang, Michael T. and Fang, Yawen and Fling, Matthew J. and Fon, Warren and Garcia, Gabriel and Gorshkov, Alexey V. and Grant, Julia A. and Gray, Mason J. and Grimberg, Sebastian and Grimsmo, Arne L. and Haim, Arbel and Hand, Justin and He, Yuan and Hernandez, Mike and Hover, David and Hung, Jimmy S. C. and Hunt, Matthew and Iverson, Joe and Jarrige, Ignace and Jaskula, Jean-Christophe and Jiang, Liang and Kalaee, Mahmoud and Karabalin, Rassul and Karalekas, Peter J. and Keller, Andrew J. and Khalajhedayati, Amirhossein and Kubica, Aleksander and Lee, Hanho and Leroux, Catherine and Lieu, Simon and Ly, Victor and Madrigal, Keven Villegas and Marcaud, Guillaume and McCabe, Gavin and Miles, Cody and Milsted, Ashley and Minguzzi, Joaquin and Mishra, Anurag and Mukherjee, Biswaroop and Naghiloo, Mahdi and Oblepias, Eric and Ortuno, Gerson and Pagdilao, Jason and Pancotti, Nicola and Panduro, Ashley and Paquette, JP and Park, Minje and Peairs, Gregory A. and Perello, David and Peterson, Eric C. and Ponte, Sophia and Preskill, John and Qiao, Johnson and Refael, Gil and Resnick, Rachel and Retzker, Alex and Reyna, Omar A. and Runyan, Marc and Ryan, Colm A. and Sahmoud, Abdulrahman and Sanchez, Ernesto and Sanil, Rohan and Sankar, Krishanu and Sato, Yuki and Scaffidi, Thomas and Siavoshi, Salome and Sivarajah, Prasahnt and Skogland, Trenton and Su, Chun-Ju and Swenson, Loren J. and Teo, Stephanie M. and Tomada, Astrid and Torlai, Giacomo and Wollack, E. Alex and Ye, Yufeng and Zerrudo, Jessica A. and Zhang, Kailing and Brand{\~a}o, Fernando G. S. L. and Matheny, Matthew H. and Painter, Oskar},
	date = {2025-02-01},
	doi = {10.1038/s41586-025-08642-7},
	id = {Putterman2025},
	journal = {Nature},
	number = {8052},
	pages = {927--934},
	title = {Hardware-efficient quantum error correction via concatenated bosonic qubits},
	url = {https://doi.org/10.1038/s41586-025-08642-7},
	volume = {638},
	year = {2025}}

@misc{lacroix2024,
      title={Scaling and logic in the color code on a superconducting quantum processor}, 
      author={Nathan Lacroix and Alexandre Bourassa and Francisco J. H. Heras and Lei M. Zhang and Johannes Bausch and Andrew W. Senior and Thomas Edlich and Noah Shutty and Volodymyr Sivak and Andreas Bengtsson and Matt McEwen and Oscar Higgott and Dvir Kafri and Jahan Claes and Alexis Morvan and Zijun Chen and Adam Zalcman and Sid Madhuk and Rajeev Acharya and Laleh Aghababaie Beni and Georg Aigeldinger and Ross Alcaraz and Trond I. Andersen and Markus Ansmann and Frank Arute and Kunal Arya and Abraham Asfaw and Juan Atalaya and Ryan Babbush and Brian Ballard and Joseph C. Bardin and Alexander Bilmes and Sam Blackwell and Jenna Bovaird and Dylan Bowers and Leon Brill and Michael Broughton and David A. Browne and Brett Buchea and Bob B. Buckley and Tim Burger and Brian Burkett and Nicholas Bushnell and Anthony Cabrera and Juan Campero and Hung-Shen Chang and Ben Chiaro and Liang-Ying Chih and Agnetta Y. Cleland and Josh Cogan and Roberto Collins and Paul Conner and William Courtney and Alexander L. Crook and Ben Curtin and Sayan Das and Sean Demura and Laura De Lorenzo and Agustin Di Paolo and Paul Donohoe and Ilya Drozdov and Andrew Dunsworth and Alec Eickbusch and Aviv Moshe Elbag and Mahmoud Elzouka and Catherine Erickson and Vinicius S. Ferreira and Leslie Flores Burgos and Ebrahim Forati and Austin G. Fowler and Brooks Foxen and Suhas Ganjam and Gonzalo Garcia and Robert Gasca and Élie Genois and William Giang and Dar Gilboa and Raja Gosula and Alejandro Grajales Dau and Dietrich Graumann and Alex Greene and Jonathan A. Gross and Tan Ha and Steve Habegger and Monica Hansen and Matthew P. Harrigan and Sean D. Harrington and Stephen Heslin and Paula Heu and Reno Hiltermann and Jeremy Hilton and Sabrina Hong and Hsin-Yuan Huang and Ashley Huff and William J. Huggins and Evan Jeffrey and Zhang Jiang and Xiaoxuan Jin and Chaitali Joshi and Pavol Juhas and Andreas Kabel and Hui Kang and Amir H. Karamlou and Kostyantyn Kechedzhi and Trupti Khaire and Tanuj Khattar and Mostafa Khezri and Seon Kim and Paul V. Klimov and Bryce Kobrin and Alexander N. Korotkov and Fedor Kostritsa and John Mark Kreikebaum and Vladislav D. Kurilovich and David Landhuis and Tiano Lange-Dei and Brandon W. Langley and Pavel Laptev and Kim-Ming Lau and Justin Ledford and Kenny Lee and Brian J. Lester and Loïck Le Guevel and Wing Yan Li and Yin Li and Alexander T. Lill and William P. Livingston and Aditya Locharla and Erik Lucero and Daniel Lundahl and Aaron Lunt and Ashley Maloney and Salvatore Mandrà and Leigh S. Martin and Orion Martin and Cameron Maxfield and Jarrod R. McClean and Seneca Meeks and Anthony Megrant and Kevin C. Miao and Reza Molavi and Sebastian Molina and Shirin Montazeri and Ramis Movassagh and Charles Neill and Michael Newman and Anthony Nguyen and Murray Nguyen and Chia-Hung Ni and Murphy Y. Niu and Logan Oas and William D. Oliver and Raymond Orosco and Kristoffer Ottosson and Alex Pizzuto and Rebecca Potter and Orion Pritchard and Chris Quintana and Ganesh Ramachandran and Matthew J. Reagor and Rachel Resnick and David M. Rhodes and Gabrielle Roberts and Eliott Rosenberg and Emma Rosenfeld and Elizabeth Rossi and Pedram Roushan and Kannan Sankaragomathi and Henry F. Schurkus and Michael J. Shearn and Aaron Shorter and Vladimir Shvarts and Spencer Small and W. Clarke Smith and Sofia Springer and George Sterling and Jordan Suchard and Aaron Szasz and Alex Sztein and Douglas Thor and Eifu Tomita and Alfredo Torres and M. Mert Torunbalci and Abeer Vaishnav and Justin Vargas and Sergey Vdovichev and Guifre Vidal and Catherine Vollgraff Heidweiller and Steven Waltman and Jonathan Waltz and Shannon X. Wang and Brayden Ware and Travis Weidel and Theodore White and Kristi Wong and Bryan W. K. Woo and Maddy Woodson and Cheng Xing and Z. Jamie Yao and Ping Yeh and Bicheng Ying and Juhwan Yoo and Noureldin Yosri and Grayson Young and Yaxing Zhang and Ningfeng Zhu and Nicholas Zobrist and Hartmut Neven and Pushmeet Kohli and Alex Davies and Sergio Boixo and Julian Kelly and Cody Jones and Craig Gidney and Kevin J. Satzinger},
      year={2024},
      eprint={2412.14256},
      archivePrefix={arXiv},
      primaryClass={quant-ph},
      url={https://arxiv.org/abs/2412.14256}, 
}

@article{Chen2024,
  doi = {10.22331/q-2024-11-07-1516},
  url = {https://doi.org/10.22331/q-2024-11-07-1516},
  title = {Benchmarking a trapped-ion quantum computer with 30 qubits},
  author = {Chen, Jwo-Sy and Nielsen, Erik and Ebert, Matthew and Inlek, Volkan and Wright, Kenneth and Chaplin, Vandiver and Maksymov, Andrii and P{\'{a}}ez, Eduardo and Poudel, Amrit and Maunz, Peter and Gamble, John},
  journal = {{Quantum}},
  issn = {2521-327X},
  publisher = {{Verein zur F{\"{o}}rderung des Open Access Publizierens in den Quantenwissenschaften}},
  volume = {8},
  pages = {1516},
  month = nov,
  year = {2024}
}

@article{
RyanAnderson2024,
author = {C. Ryan-Anderson  and N. C. Brown  and C. H. Baldwin  and J. M. Dreiling  and C. Foltz  and J. P. Gaebler  and T. M. Gatterman  and N. Hewitt  and C. Holliman  and C. V. Horst  and J. Johansen  and D. Lucchetti  and T. Mengle  and M. Matheny  and Y. Matsuoka  and K. Mayer  and M. Mills  and S. A. Moses  and B. Neyenhuis  and J. Pino  and P. Siegfried  and R. P. Stutz  and J. Walker  and D. Hayes },
title = {High-fidelity teleportation of a logical qubit using transversal gates and lattice surgery},
journal = {Science},
volume = {385},
number = {6715},
pages = {1327-1331},
year = {2024},
doi = {10.1126/science.adp6016},
URL = {https://www.science.org/doi/abs/10.1126/science.adp6016}}

@article{Egan2021,
	author = {Egan, Laird and Debroy, Dripto M. and Noel, Crystal and Risinger, Andrew and Zhu, Daiwei and Biswas, Debopriyo and Newman, Michael and Li, Muyuan and Brown, Kenneth R. and Cetina, Marko and Monroe, Christopher},
	date = {2021-10-01},
	doi = {10.1038/s41586-021-03928-y},
	id = {Egan2021},
	journal = {Nature},
	number = {7880},
	pages = {281--286},
	title = {Fault-tolerant control of an error-corrected qubit},
	url = {https://doi.org/10.1038/s41586-021-03928-y},
	volume = {598},
	year = {2021}}

@misc{rodriguez2024,
      title={Experimental Demonstration of Logical Magic State Distillation}, 
      author={Pedro Sales Rodriguez and John M. Robinson and Paul Niklas Jepsen and Zhiyang He and Casey Duckering and Chen Zhao and Kai-Hsin Wu and Joseph Campo and Kevin Bagnall and Minho Kwon and Thomas Karolyshyn and Phillip Weinberg and Madelyn Cain and Simon J. Evered and Alexandra A. Geim and Marcin Kalinowski and Sophie H. Li and Tom Manovitz and Jesse Amato-Grill and James I. Basham and Liane Bernstein and Boris Braverman and Alexei Bylinskii and Adam Choukri and Robert DeAngelo and Fang Fang and Connor Fieweger and Paige Frederick and David Haines and Majd Hamdan and Julian Hammett and Ning Hsu and Ming-Guang Hu and Florian Huber and Ningyuan Jia and Dhruv Kedar and Milan Kornjača and Fangli Liu and John Long and Jonathan Lopatin and Pedro L. S. Lopes and Xiu-Zhe Luo and Tommaso Macrì and Ognjen Marković and Luis A. Martínez-Martínez and Xianmei Meng and Stefan Ostermann and Evgeny Ostroumov and David Paquette and Zexuan Qiang and Vadim Shofman and Anshuman Singh and Manuj Singh and Nandan Sinha and Henry Thoreen and Noel Wan and Yiping Wang and Daniel Waxman-Lenz and Tak Wong and Jonathan Wurtz and Andrii Zhdanov and Laurent Zheng and Markus Greiner and Alexander Keesling and Nathan Gemelke and Vladan Vuletić and Takuya Kitagawa and Sheng-Tao Wang and Dolev Bluvstein and Mikhail D. Lukin and Alexander Lukin and Hengyun Zhou and Sergio H. Cantú},
      year={2024},
      eprint={2412.15165},
      archivePrefix={arXiv},
      primaryClass={quant-ph},
      url={https://arxiv.org/abs/2412.15165}, 
}

@misc{reichardt2024,
      title={Logical computation demonstrated with a neutral atom quantum processor}, 
      author={Ben W. Reichardt and Adam Paetznick and David Aasen and Ivan Basov and Juan M. Bello-Rivas and Parsa Bonderson and Rui Chao and Wim van Dam and Matthew B. Hastings and Andres Paz and Marcus P. da Silva and Aarthi Sundaram and Krysta M. Svore and Alexander Vaschillo and Zhenghan Wang and Matt Zanner and William B. Cairncross and Cheng-An Chen and Daniel Crow and Hyosub Kim and Jonathan M. Kindem and Jonathan King and Michael McDonald and Matthew A. Norcia and Albert Ryou and Mark Stone and Laura Wadleigh and Katrina Barnes and Peter Battaglino and Thomas C. Bohdanowicz and Graham Booth and Andrew Brown and Mark O. Brown and Kayleigh Cassella and Robin Coxe and Jeffrey M. Epstein and Max Feldkamp and Christopher Griger and Eli Halperin and Andre Heinz and Frederic Hummel and Matthew Jaffe and Antonia M. W. Jones and Eliot Kapit and Krish Kotru and Joseph Lauigan and Ming Li and Jan Marjanovic and Eli Megidish and Matthew Meredith and Ryan Morshead and Juan A. Muniz and Sandeep Narayanaswami and Ciro Nishiguchi and Timothy Paule and Kelly A. Pawlak and Kristen L. Pudenz and David Rodríguez Pérez and Jon Simon and Aaron Smull and Daniel Stack and Miroslav Urbanek and René J. M. van de Veerdonk and Zachary Vendeiro and Robert T. Weverka and Thomas Wilkason and Tsung-Yao Wu and Xin Xie and Evan Zalys-Geller and Xiaogang Zhang and Benjamin J. Bloom},
      year={2024},
      eprint={2411.11822},
      archivePrefix={arXiv},
      primaryClass={quant-ph},
      url={https://arxiv.org/abs/2411.11822}, 
}

@article{gidney2025factor,
doi = {10.22331/q-2021-04-15-433},
url = {https://doi.org/10.22331/q-2021-04-15-433},
title = {How to factor 2048 bit {RSA} integers in 8 hours using 20 million noisy qubits},
author = {Gidney, Craig and Eker{\aa{}}, Martin},
journal = {{Quantum}},
issn = {2521-327X},
publisher = {{Verein zur F{\"{o}}rderung des Open Access Publizierens in den Quantenwissenschaften}},
volume = {5},
pages = {433},
month = apr,
year = {2021}
}

@article{McArdle2020,
  title = {Quantum computational chemistry},
  author = {McArdle, Sam and Endo, Suguru and Aspuru-Guzik, Al\'an and Benjamin, Simon C. and Yuan, Xiao},
  journal = {Rev. Mod. Phys.},
  volume = {92},
  issue = {1},
  pages = {015003},
  numpages = {51},
  year = {2020},
  month = {3},
  publisher = {American Physical Society},
  doi = {10.1103/RevModPhys.92.015003},
  url = {https://link.aps.org/doi/10.1103/RevModPhys.92.015003}
}

@article{zhong2019violating,
  title={Violating Bell’s inequality with remotely connected superconducting qubits},
  author={Zhong, YP and Chang, H-S and Satzinger, KJ and Chou, M-H and Bienfait, Audrey and Conner, CR and Dumur, {\'E} and Grebel, Joel and Peairs, GA and Povey, RG and others},
  journal={Nature Physics},
  volume={15},
  number={8},
  pages={741--744},
  year={2019},
  publisher={Nature Publishing Group UK London}
}

@article{Alexeev2024,
   title={Quantum-centric supercomputing for materials science: A perspective on challenges and future directions},
   volume={160},
   ISSN={0167-739X},
   url={http://dx.doi.org/10.1016/j.future.2024.04.060},
   DOI={10.1016/j.future.2024.04.060},
   journal={Future Generation Computer Systems},
   publisher={Elsevier BV},
   author={Alexeev, Yuri and Amsler, Maximilian and Barroca, Marco Antonio and Bassini, Sanzio and Battelle, Torey and Camps, Daan and Casanova, David and Choi, Young Jay and Chong, Frederic T. and Chung, Charles and Codella, Christopher and Córcoles, Antonio D. and Cruise, James and Di Meglio, Alberto and Duran, Ivan and Eckl, Thomas and Economou, Sophia and Eidenbenz, Stephan and Elmegreen, Bruce and Fare, Clyde and Faro, Ismael and Fernández, Cristina Sanz and Ferreira, Rodrigo Neumann Barros and Fuji, Keisuke and Fuller, Bryce and Gagliardi, Laura and Galli, Giulia and Glick, Jennifer R. and Gobbi, Isacco and Gokhale, Pranav and de la Puente Gonzalez, Salvador and Greiner, Johannes and Gropp, Bill and Grossi, Michele and Gull, Emanuel and Healy, Burns and Hermes, Matthew R. and Huang, Benchen and Humble, Travis S. and Ito, Nobuyasu and Izmaylov, Artur F. and Javadi-Abhari, Ali and Jennewein, Douglas and Jha, Shantenu and Jiang, Liang and Jones, Barbara and de Jong, Wibe Albert and Jurcevic, Petar and Kirby, William and Kister, Stefan and Kitagawa, Masahiro and Klassen, Joel and Klymko, Katherine and Koh, Kwangwon and Kondo, Masaaki and Kürkçüog̃lu, Dog̃a Murat and Kurowski, Krzysztof and Laino, Teodoro and Landfield, Ryan and Leininger, Matt and Leyton-Ortega, Vicente and Li, Ang and Lin, Meifeng and Liu, Junyu and Lorente, Nicolas and Luckow, Andre and Martiel, Simon and Martin-Fernandez, Francisco and Martonosi, Margaret and Marvinney, Claire and Medina, Arcesio Castaneda and Merten, Dirk and Mezzacapo, Antonio and Michielsen, Kristel and Mitra, Abhishek and Mittal, Tushar and Moon, Kyungsun and Moore, Joel and Mostame, Sarah and Motta, Mario and Na, Young-Hye and Nam, Yunseong and Narang, Prineha and Ohnishi, Yu-ya and Ottaviani, Daniele and Otten, Matthew and Pakin, Scott and Pascuzzi, Vincent R. and Pednault, Edwin and Piontek, Tomasz and Pitera, Jed and Rall, Patrick and Ravi, Gokul Subramanian and Robertson, Niall and Rossi, Matteo A.C. and Rydlichowski, Piotr and Ryu, Hoon and Samsonidze, Georgy and Sato, Mitsuhisa and Saurabh, Nishant and Sharma, Vidushi and Sharma, Kunal and Shin, Soyoung and Slessman, George and Steiner, Mathias and Sitdikov, Iskandar and Suh, In-Saeng and Switzer, Eric D. and Tang, Wei and Thompson, Joel and Todo, Synge and Tran, Minh C. and Trenev, Dimitar and Trott, Christian and Tseng, Huan-Hsin and Tubman, Norm M. and Tureci, Esin and Valiñas, David García and Vallecorsa, Sofia and Wever, Christopher and Wojciechowski, Konrad and Wu, Xiaodi and Yoo, Shinjae and Yoshioka, Nobuyuki and Yu, Victor Wen-zhe and Yunoki, Seiji and Zhuk, Sergiy and Zubarev, Dmitry},
   year={2024},
   month=nov, pages={666–710} }

@inproceedings{huan2025towards,
  title={Towards Quantum Repeater Nodes with Weakly-Coupled Nuclear Spins in Diamond},
  author={Huan, Yan Qi and Suleymanzade, Aziza and Stas, Pieter-Jan and Wei, Yan-Cheng and Knall, Erik N and Arias, Francisca Abdo and Kniazev, Evgenii and Machielse, Bart and Yazlar, Umut and Baranes, Gefen and others},
  booktitle={Quantum 2.0},
  pages={QW4A--3},
  year={2025},
  organization={Optica Publishing Group}
}

\clearpage
\appendix
\numberwithin{equation}{section}

\section{Balanced pipeline for entanglement distillation} \label{app:distillation rate}
Here we derive the formulae needed for computing the distillation rate of a given distillation sequence.

In steady state, the optimal pipeline is such that instances of each stage of the sequence are run as soon as the required inputs have been prepared. This means that each stage of the sequence only ever waits because of insufficient inputs and not because of insufficient memory. This is called the balanced pipeline. In many cases, ours included, the output rate $r_\mathrm{out}$ as well as the memory average usage of actively running processes $M^\mathrm{active}_S$ of a balanced pipeline are directly proportional to the input rate. Taking into account the average space taken up by idling input qubits $M^\mathrm{idle}_S$ we have
\begin{gather}
    r_\mathrm{out} = E_S r_\mathrm{in}
\intertext{and}
    M_\mathrm{total} = M^\mathrm{idle}_S + M^\mathrm{active}_S = M^\mathrm{idle}_S + M_S r_\mathrm{in},
\end{gather}
where $E_S$, $ M^\mathrm{idle}_S$, and $M_S$ are constants determined by the details of the given sequence $S$. If the available memory is limited, we find that we must choose
\begin{equation}
    r_\mathrm{in} \leq \frac{M_\mathrm{total} - M^\mathrm{idle}_S}{M_S} \define C_S.
\end{equation}
During operations, random fluctuations will mean that the space taken up by the instances of a stage might be larger than the average, but simultaneously others will take up less. This effect in addition to the buffer that is made available through $M^\mathrm{idle}_S$, makes cases where the balanced pipeline will hit the memory cap exceedingly rare. Thus, to a good approximation we can take equality in the above equation and can still expect the output rate to be given by $r_\mathrm{out} = E_S r_\mathrm{in}$. This we also confirm empirically by Monte Carlo simulation (see \cite{marqversen_quantum_computations_2025}).

Obviously the input rate cannot exceed the rate at which physical Bell pairs can be provided $r_\mathrm{bell}$. So we get two regimes: Input limited and memory limited:
\begin{equation}
    r_\mathrm{out} = E_S r_\mathrm{in} = E_S \min(C_S, r_\mathrm{bell}).
\end{equation}
This is the general result presented in the main text.

In the remainder of this section we derive the specific formulae for $E_S$, $M_S$, and $M^\mathrm{idle}_S$ given an arbitrary distillation sequence $S$. A balanced pipeline should essentially be viewed as a fully parallelised process with no memory constraint. This is exactly the regime analysed in the appendices of the original work on entanglement distillation \cite{pattison_fast_2024}. We will in large use the notation and results from there.

A distillation sequence $S$ is a sequence of stages $(S_i)$. Each stage, $S_i$, that being code distillation, growing, or injection, can be described by the following parameters: The code parameters $[[n_i, k_i, d_i]]$; the rejection rate $p^\mathrm{fail}_i$ which in general depends on, and is thus calculated from, the output error of the previous stage $p^\mathrm{out}_{i-1}$; the execution time $T_i$; and the logical qubit size $s_i$ in terms of physical qubits.

At each stage $i$ of a sequence $S$ we define the encoding rate $E_i$ which describe the expected ratio of logical Bell pairs output from stage $i$ to initial physical Bell pairs, and the number $K_i$ of logical qubits that the stage outputs per successful distillation attempt:
\begin{equation}
    E_i = \prod_{j=0}^i (1 - p^\mathrm{fail}_j) \frac{k_j}{n_j}
        \qq{and}
    K_i = \prod_{j=0}^i k_i.
\end{equation}
Since instances are started as soon as enough inputs are prepared, this means that the output rate of stage $i$ is $r^\mathrm{out}_i = (1 - p^\mathrm{fail}_i) \frac{k_i}{n_i} r^\mathrm{out}_{i-1}$. This immediately leads to
\begin{equation}
    r_\mathrm{out} = E_S r_\mathrm{in}
        \qq{where}
    E_S = \prod_i (1 - p^\mathrm{fail}_i) \frac{k_i}{n_i},
\end{equation}
with $E_S$ being the encoding rate of the complete sequence.

The average number of instances of stage $i$ that will be running in parallel is given by
\begin{equation}
    N_i = T_i r^{init}_i = T_i \frac{K_{i-1}}{n_i} E_{i-1} r_\mathrm{in},
\end{equation}
where $r^{init}_i$ is the rate at which inputs are prepared, and $r_\mathrm{in}$ is the rate at which physical Bell pairs are supplied at the start of the whole sequence. Each instance takes up $n_i$ qubits of size $s_i$, and so the average space taken up by active instances of stage $i$ is
\begin{equation}
    M^\mathrm{active}_i = N_i n_i s_i = s_i T_i E_{i-1} K_{i-1} r_\mathrm{in}.
\end{equation}
Since instances are initialised as soon as inputs are ready, the average space taken up by idle input qubits to stage $i$ is
\begin{equation}
    M^\mathrm{idle}_i = \half n_i s_i K_{i-1}.
\end{equation}
Summing up the contributions of all stages gives
\begin{align}
        & & 
    M_\mathrm{total} &= M^\mathrm{idle}_S + M_S r_\mathrm{in}
        \\ &\text{with} \quad &
    M^\mathrm{idle}_S &= \sum_i \half n_i s_i K_{i-1}
        \\ &\text{and} \quad &
    M_S &= \sum_i s_i T_i E_{i-1} K_{i-1}.
\end{align}
These are used for explicitly computing $C_S$ which then is used to evaluate the distillation rate.

\section{Distillation sequence optimisation} \label{app:sequence optimisation}

\subsection{DFS Branch pruning}
The objective function that we wish to maximise is the distillation rate:
\begin{equation}
    r^\mathrm{out}_S = \min(E_S C_S, E_S r_\mathrm{bell}),
\end{equation}
which should be maximised over the space of valid distillation sequences $S$. This can be done by a standard depth-first search (DFS). However, the space of sequences is huge and has a very large branching factor. Clever branch pruning is thus required to make the search feasible. Here we present the details of our strategy.

Let $S$ and $S'$ be two distillation sequences. Let $S + S'$ denote the sequence that is the concatenation starting with $S$ and followed by $S'$. We will consider the objective function evaluated for the concatenated sequence $r^\mathrm{out}_{S + S'}$. Using the definitions and results from \cref{app:distillation rate} we note the following relations
\begin{gather}
    E_{S + S'} = E_S \cdot E_{S'}(p^\mathrm{out}_S, L_S)
        \intertext{and}
    C_{S + S'} = \frac{M - M^\mathrm{idle}_S - K_S M^\mathrm{idle}_{S'}}{M_S + E_S K_S M_{S'}(p^\mathrm{out}_S, L_S)},
\end{gather}
where $p^\mathrm{out}_S$ and $L_S$ is the error rate and code size of the logical Bell-pairs output from sequence $S$ respectively. Both of these are included explicitly since the success rates and distilled error rate of the stages of $S'$ depend on these values. $M^\mathrm{idle}_{S'}$ however is independent of $S$. These relations identify the fundamentally underlying parameters dictating the value of the objective function.
\begin{align}
    r^\mathrm{out}_{S + S'} &= r_{S'}(M^\mathrm{idle}_S, M_S, K_S, E_S, p^\mathrm{out}_S, L_S) 
        \\ &\define
    r_{S'}(\vb{v}_S, L_S),
\end{align}
where $\vb{v}_S$ is the vector of parameters excluding $L_S$. It turns out (to be shown) that $r_{S'}$ is a monotonous function in each of the parameters of $\vb{v}_S$ independent of the extension $S'$.

To see how the above observation is relevant, we assume, without loss of generality, that the rate is decreasing in each of the parameters. Now consider any two sequences $S$ and $S'$ with $L_S = L_{S'}$ and $\text{v}_{S',i} \geq \text{v}_{S,i}$ for all $i$. Since the objective function is decreasing we can immediately conclude that for any extension $S''$ we have
\begin{equation}
    r^\mathrm{out}_{S' + S''} \leq r^\mathrm{out}_{S + S''},
\end{equation}
% \textcolor{red}{Maybe something about only comparing sequences that are also solutions...}
% It is not really helpful to be comparing the objective function of sequences that are not also \emph{solutions}. By solution we mean a distillation sequence that outputs states with error rate $p^\mathrm{out}_S < p_{target}$. Assume that we have previously explored all branches extending from $S$. Then, since $L_S = L_{S'}$ and $p^\mathrm{out}_S \leq p^\mathrm{out}_{S'}$...
and so, the branch extending from $S'$ can be cut immediately without exploring it any further.

What is left is to show is that $r_{S'}$ is monotonous in each of the parameters of $\vb{v}$. From the explicit form of $C_{S+S'}$ this is trivially true for the parameters $M^\mathrm{idle}_S$, $M_S$, and $K_S$. Clearly the input limited case $E_S r_\mathrm{bell}$ is increasing in $E_S$. This also holds for the memory limited case $E_S C_S$ since
\begin{equation}
    \pdv{}{E_S} \frac{E_S}{M_S + E_S K_S M_{S'}} 
        =
    \frac{M_S}{(M_S + E_S K_S M_{S'})^2}
        \geq 
    0.
\end{equation}
Finally we will show that the rate is decreasing in $p^\mathrm{out}_S$. From the definition of $E_i$ it follows
\begin{equation}
    \pdv{E_i}{p^\mathrm{out}_S} = - A_i E_i
        \qq{with}
    0 \leq A_i \leq A_{i+1}.
\end{equation}
This immediately covers the input limited case. From the definition of $M_{S'}$ and using the above result we get
\begin{align}
    \pdv{M_{S'}}{p^\mathrm{out}_S} 
        &=
    \pdv{}{p^\mathrm{out}_S} \sum_j s_i T_i K_{i-1} E_{i-1}
        \\ &=
    - \sum_j s_i T_i K_{i-1} E_{i-1} A_{i-1}
        \\ &\geq
    - \sum_j s_i T_i K_{i-1} E_{i-1} A_{S'}
        \\ &=
    - A_{S'} M_{S'}.
\end{align}
Putting this inequality to use we find
\begin{gather}
    \pdv{}{p^\mathrm{out}_S} (E_{S+S'} C_{S+S'})
        \\ =
    \abs{B} \Big[ \partial E_{S'} (M_S + E_S K_S M_{S'}) - E_{S'} E_S K_S \partial M_{S'} \Big]
        \\ =
    - \abs{B} \Big[ A_{S'} E_{S'} M_S + E_{S'} E_S K_S (A_{S'} M_{S'} + \partial M_{S'}) \Big]
        \\ \leq
    - \abs{B} A_{S'} E_{S'} M_S
        \leq
    0.
\end{gather}
We thus conclude that the objective function indeed is decreasing in $p^\mathrm{out}_S$.

\subsection{Reducing the search space}
With the pruning method described in the previous section, the DFS is feasible and somewhat fast. However, to explore the 2D parameter space of \cref{fig:top view} require a huge number of separate optimisations.

Since the distillation rate of a sequence depends in a non-trivial way on the physical Bell pair rate $r_\mathrm{bell}$, so does the optimal sequence
\begin{equation}
    S^*_\lambda 
    = \argmax_S[r^\mathrm{out}_S(\lambda)] 
    = \argmax_S \left[ \min(E_S \lambda, E_S C_S) \right]
\end{equation}
where we from now on use the simplifying notation
\begin{equation}
    \lambda = r_\mathrm{bell}.
\end{equation}
And so, naively each single point in \cref{fig:top view} require a separate search. We now discuss how this number can be reduced as to get exact values for the vast majority of the space, while also providing a quite reasonable lower bound in the remainder.

Indeed, for a fixed memory constraint, there are two sequences that in general are optimal for a large range of rates. These sequences are the sequence with the largest encoding rate and the sequence with the largest distillation rate cap:
\begin{align}
    S' &= \argmax_S \left[ E_S \right]
\intertext{and}
    S'' &= \argmax_S \left[ E_S C_S \right].
\end{align}
In the limit of low Bell rates $S'$ will always be optimal, and in the limit of high Bell rates $S''$ will be optimal. More precisely
\begin{align}
    \lambda \leq C_{S'} &\implies S^* = S'
\intertext{and}
    \lambda \geq C_{S''} &\implies S^* = S''.
\end{align}
These statements follow directly from the definition of $r^\mathrm{out}_S(\lambda)$, $S'$, and $S''$. The intuition is that when sequences are not memory limited, the distillation rate is linear in the Bell rate, and so, the sequence with the largest encoding rate $S'$ always is optimal. However, as soon as $S'$ becomes memory limited, $\lambda > C_{S'}$, the corresponding distillation rate caps out at $E_{S'} C_{S'}$ which is suboptimal. Equivalently, no sequence can ever provide a higher distillation rate than $E_{S''} C_{S''}$, so as soon as $S''$ reaches that rate $\lambda \geq C_{S''}$, no other sequence can beat that. 

\begin{figure}
    % \makebox[\linewidth][c]{
    %     \includegraphics[width=0.6\linewidth]{figures/reduced_optimisation.png}
    %     \hspace{5mm}
    %     \includegraphics[width=0.6\linewidth]{figures/complete_optimisation.png}
    % }%
    \centering
    \includegraphics[width=\linewidth]{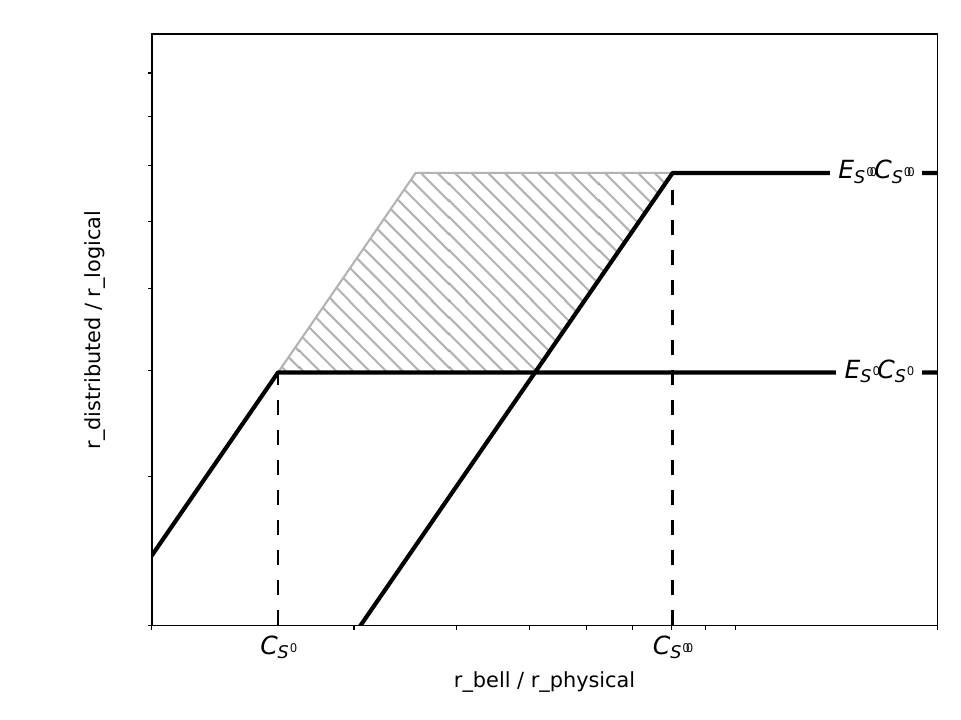}
    \includegraphics[width=\linewidth]{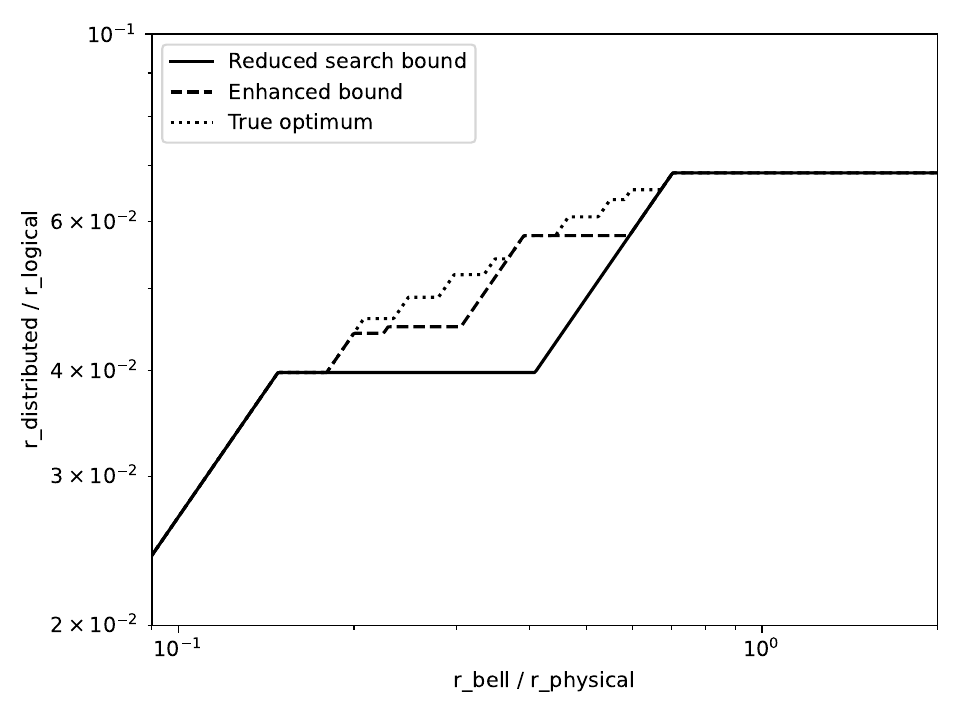}
    \caption{(Left) The two solid lines illustrate the rates produced by the optimal sequences in the limit of low $S'$ and high Bell pair rate $S''$. Together, they provide the maximal distillation rate for most cases. Only for input rates between $C_{S'}$ and $C_{S''}$ can sequences exist that further improve on these rates. Furthermore, the optimum is known to be restricted to within the gray shaded area. (Right) The distillation rate for allocated space of 13565 physical qubits and target error $p_{target}=10^{-12}$. The solid line is the bound given by the reduced search. The dashed line is the enhanced bound induced by solutions for space <13565, and the dotted line is the true maximum rate.}
    \label{fig:reduced space optimisation}
\end{figure}

Only when $\lambda$ is between $C_{S'}$ and $C_{S''}$ do these two sequences not provide the optimal distillation rate. However, they do provide lower bounds on the optimum which significantly restrict the possibilities. Specificity the optimal distillation rate $r^*_\lambda$ must lie in a small well-defined region. Specifically when
\begin{gather}
    \lambda \in (C_{S'}, C_{S''})
\intertext{then}
    \min(E_{S'} C_{S'}, E_{S''} \lambda) \leq r^*_\lambda \leq \min(E_{S'} \lambda, E_{S''} C_{S''}).
\end{gather}
All of these ideas are illustrated in \cref{fig:reduced space optimisation}. By restricting the search to only find $S'$ and $S''$ we gain a really good bound on the actual optimum across all values of $\lambda$.

In making \cref{fig:top view} we further improve on the lower bound provided by the two limiting cases, by using the fact that a sequence that successfully distils Bell pairs at allocated memory $M$ also is a valid sequence for all cases $\geq M$. Since ``good'' sequences for one value of $M$ can be expected to be good sequences for values in the neighbourhood, this strategy effectively enhances the known bound to be close to optimal as is evident from \cref{fig:reduced space optimisation}. Upon applying this strategy, we find that the number of \emph{unique} sequences that are relevant is only on the order of 10s of sequences. We therefore get close to optimal results by finding the optimal sequences for only a very modest number of distinct $M$s, reducing the amount of work even further.

\section{Surface Code Growing} \label{app:growing}

\begin{figure}
    \centering
    \includegraphics[width=\linewidth]{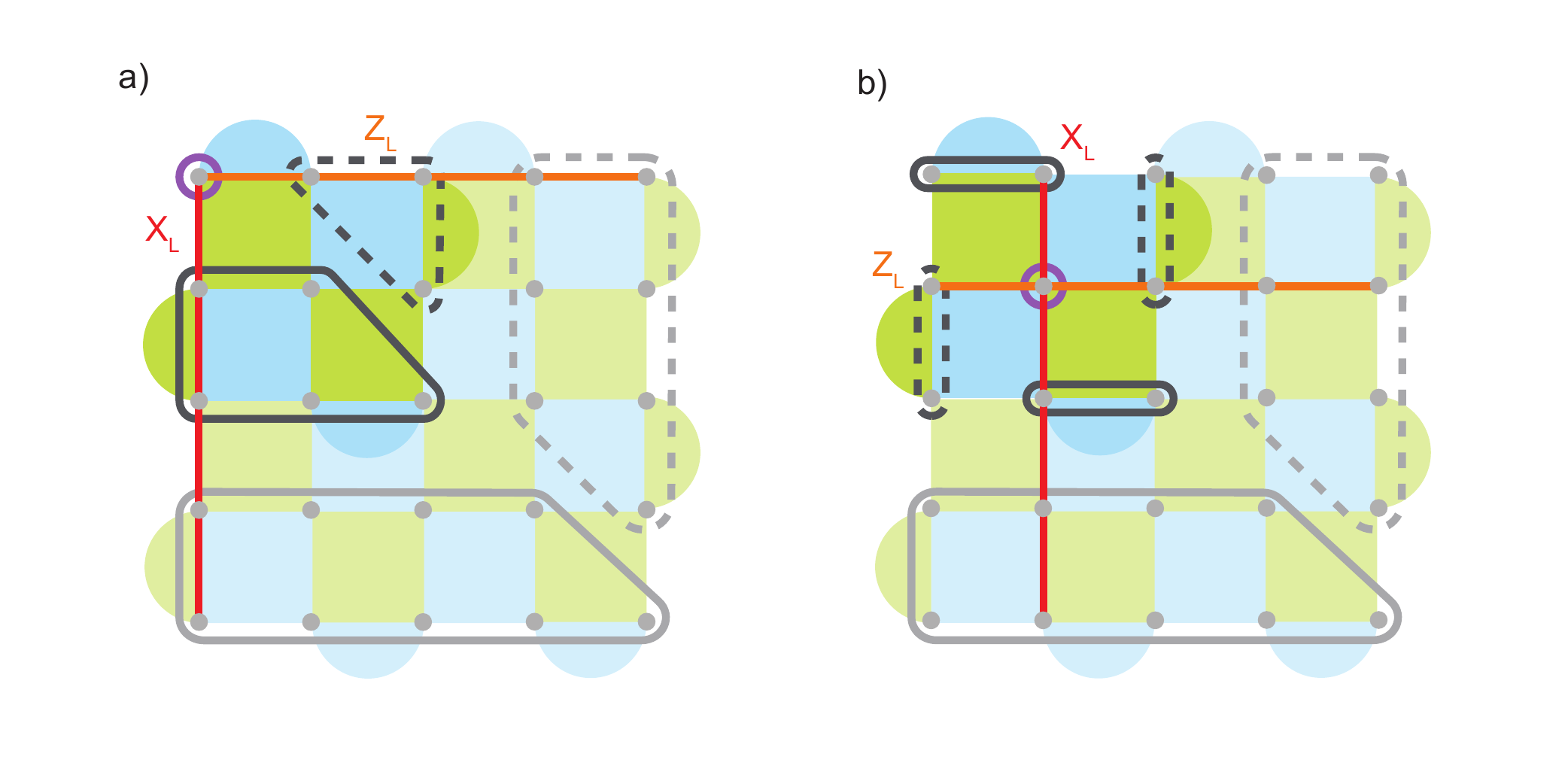}
    \caption{Schematic of rotated surface code injection and growing from $d = 3$ to $d = 5$: (a) corner injection and (b) middle injection. Data qubits enclosed by solid lines are initialised in the $\ket{+}$ state; those with dashed lines are initialised in the $\ket{0}$ state. The purple circle denotes the injection qubit (from a Bell pair). Green and blue represent $Z$ and $X$ stabilisers, respectively. Logical operators $Z_L$ and $X_L$ are shown as an orange horizontal and red vertical line, respectively.}
    \label{fig:growing_scheme}
\end{figure}

We implement a surface code growing procedure based on the method introduced in~\cite{li_magic_2015}, adapted here for the rotated surface code. This approach begins with a rotated code of distance $d_{\mathrm{start}}$ and grows it to a larger code of distance $d_{\mathrm{final}}$. To maintain the consistency of logical operators during growth, additional patches are initialised in the $\ket{+}$ state along the right edge and in the $\ket{0}$ state along the bottom edge (see \cref{fig:growing_scheme}).

Both corner injection~\cite{li_magic_2015} and middle injection~\cite{lao_magic_2022} are compatible with this growing technique. Since their logical error rates (LER) and rejection rates are comparable, we use corner injection in our simulations for simplicity. Code growth is performed by executing one round of stabiliser measurements over the full $d_{\mathrm{final}}$ lattice. Detectors are placed on stabilisers that were part of the initial code as well as on the newly added stabilisers required for the grown patch.

\begin{figure}
    \centering
    \includegraphics[width=\linewidth]{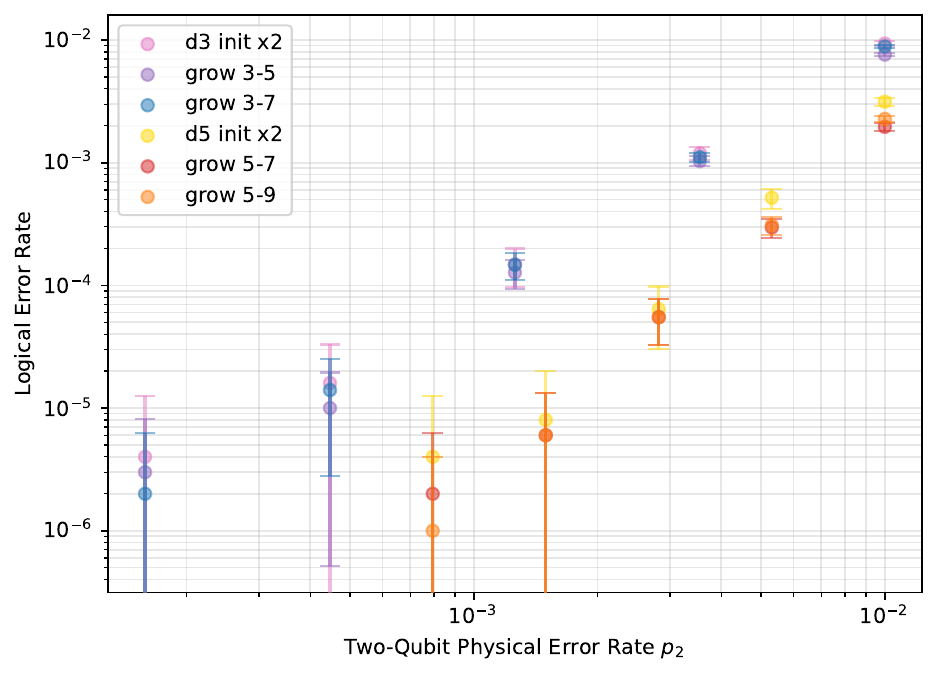}
    \caption{Comparison of logical error rate for growing versus direct initialisation of a surface code for starting distances $d_{\mathrm{start}} = 3$ and $5$.}
    \label{fig:growth scaling}
\end{figure}

We simulate this growing procedure and compare its LER to that of standard code initialisation. The results are shown in \cref{fig:growth scaling}. In our simulations, a noiseless $d_{\mathrm{start}}$ code is prepared in the logical $\ket{0}$ state using a single round of stabiliser measurements. The code is then grown to $d_{\mathrm{final}}$ using noisy patch initialisation and noisy stabiliser measurement, followed by a logical $Z_L$ measurement. We evaluate four growing transitions: $3 \rightarrow 5$, $3 \rightarrow 7$, $5 \rightarrow 7$, and $5 \rightarrow 9$. Simulations are performed under a biased circuit noise model in which single-qubit operations have an error rate $p_1 = p_2 / 10$, where $p_2$ is the error rate for two-qubit operations.

Our results show that the logical error introduced during code growing depends only on the initial code distance $d_{\mathrm{start}}$, and is approximately twice the LER of direct initialisation for a code of that same distance:
\[
p_{\text{growing: }d_{\text{start}} \rightarrow d_{\mathrm{final}} } \approx 2A_{d_{\text{start}}} \left( \frac{p_2}{p_b^*}\right)^{\left( \frac{d_{\mathrm{start}}+1}{2} \right)},
\]
where $p_b^*$ is a bulk threshold, and constant $A_{d_{\text{start}}}$ depends only on the initial code distance $d_{\mathrm{start}}$ and characterises LER of direct initialisation: 
\[
p_{\text{init }d_{\text{start}}} \approx A_{d_{\text{start}}} \left( \frac{p_2}{p_b^*}\right)^{\left( \frac{d_{\mathrm{start}}+1}{2} \right)}.
\]

For comparison, the initialisation LER is obtained by preparing a full $d_{\mathrm{start}} = 3$ or $5$ code via a single round of stabiliser measurements across the entire patch.

The simulation is performed using the Python library \texttt{Stim}~\cite{gidney2021stim} and is available online at~\cite{sirotin_growing_injection_2025}.

\section{Overview of Hardware Platforms for Distributed Quantum Computation} \label{app:exp_params}

Several physical platforms are under active development for realising distributed quantum computation. Here, we describe three prominent architectures and summarise their current experimental benchmarks and future goals.

\subsection{Solid State Defects in Nanophotonic Cavities:} 
Each network node contains a group IV defect in diamond (e.g., SiV) coupled to a nanophotonic cavity, offering strong light–matter coupling and making this a promising platform for quantum networking~\cite{stas2022robust,Cheng_SiV_BQC,knaut_entanglement_2024,bartling2024universal}.
Similar to Ref.~\cite{baranes2025designing}, we envision a platform with an ensemble of such cavities, each hosting a communication qubit (electron spin) and a long-lived memory qubit (nuclear spin).
Local deterministic gates between memory qubits are mediated via communication qubits and optical photons, using a loss-tolerant architecture~\cite{baranes2025designing}.
In this way, a distributed logical qubit can be implemented across multiple cavities, using the local photonic gates within a node.
Remote entanglement is established using heralded spin–photon gates and transferred to memory qubits for storage. 
Since both local and distributed gates are optically mediated, we assume that in the future $r_{\mathrm{Bell}} \leq r_{\mathrm{Local}}$, meaning the distributed gate rate may approach or remain slower than the local gate rate.
This platform benefits from both cavity enhancement and telecom compatibility~\cite{knaut_entanglement_2024}.

% \color{red}
% assume $r_\mathrm{bell} <= r_{local}$.
% change to group 4
% emphasize what we mean by local gates (mediated by comm qubits, electron qubits).
% distributed also with photons..
% idling error: check that today demonstrated coherence time is 1 sec and in the future 10 sec is OK
% \color{black}

\begin{table}[H]
    \scriptsize
    \centering
    \begin{tabular}{|l|l|l|}
    \hline
    \textbf{Parameter} & \textbf{Demonstrated} & \textbf{Future Goal} \\
    \hline
    Local gate error & $\sim$7\% \cite{bartling2024universal,stas2022robust,Cheng_SiV_BQC} & 0.1\% \\
    Local gate time & $\sim$50$~\mu$s \cite{Cheng_SiV_BQC} & 1~$\mu$s \\
    Distr. gate error & $\sim$10\% \cite{knaut_entanglement_2024, Cheng_SiV_BQC} & 1\% \\
    Distr. gate rate & 1~Hz \cite{knaut_entanglement_2024} & 100~kHz -- 1~MHz \\
    Idling error rate &  0.1\% per ms \cite{Cheng_SiV_BQC, knaut_entanglement_2024, bartling2024universal} & $10^{-4}$ per ms \\
    Memory size & 3 qubits \cite{huan2025towards, Cheng_SiV_BQC} & 100 -- 5\,000 \\
    \hline
    \end{tabular}
    \caption{Group IV defects in diamonds platform parameters.  Each node uses an electron spin for communication and a nuclear spin for storage. “Local” refers to on-chip operations delegated by optical photons, while “Distributed” parameters refer to entanglement between nodes via photonic links.}
\end{table}

\subsection{Superconducting Qubits:}
The superconducting platform holds strong promise for quantum computing, with each node comprising multiple chips in a cryogenic environment and inter-node links implemented using photons~\cite{majidy2024building}.
Local gates between superconducting qubits on the same chip achieve high fidelities and sub-100\,ns operation times~\cite{google_2024_below_threshold_SC}.
Future architectures are expected to use microwave links between chips within a fridge, with slightly higher error tolerance: local gates may operate at 0.1\% error with a 1\% threshold, whereas inter-chip gates may allow 1\% error with a 10\% threshold~\cite{sinclair_2025neutral_atoms_interconnect}.
Distributed entanglement generation between fridges is expected to rely on microwave photons or transduction to optical~\cite{Wallraff_SC_entanglement_BellTest,Krastanov_PRL_2021_SC_modular}.
Unlike the previous platforms, this architecture performs logical gates using lattice surgery, not transversally.
As a result, logical two-qubit operations require $O(d)$ stabiliser measurements rather than $O(1)$, increasing the time overhead for distributed computation~\cite{cain_correlated_2024,baranes2025leveraging,zhou_algorithmic_2024,cain_fast_2025}.

\begin{table}[H]
    \scriptsize
    \centering
    \begin{tabular}{|l|l|l|}
    \hline
    \textbf{Parameter} & \textbf{Demonstrated} & \textbf{Future Goal} \\
    \hline
    Local gate error & $\sim$0.5\% \cite{acharya_suppressing_2023,google_2024_below_threshold_SC} & 0.1\% \\
    Local gate time & 20 ns \cite{acharya_suppressing_2023,google_2024_below_threshold_SC} & $\sim$1 ns \\
    Distr. gate error & $\sim$5\% \cite{zhong2019violating} & 1\% \\
    Distr. gate rate & $>1$~MHz \cite{zhong2019violating} & 1 -- 10~MHz \\
    Idling error rate & 1\% per $\mu$s \cite{acharya_suppressing_2023,google_2024_below_threshold_SC} & 0.1\% per $\mu$s \\
    Memory size & $\sim 100$ \cite{acharya_suppressing_2023,google_2024_below_threshold_SC} &  1\,000-10\,000 \\
    \hline
    \end{tabular}
    \caption{Superconducting platform parameters. While local operations are extremely fast and high-fidelity on superconducting qubits, the microwave links between chips or between nodes currently imposes a bottleneck. Future improvements are required to realise high-rate distributed gates with superconducting processors.}
\end{table}

\subsection{Neutral Atoms with Optical Cavities:}
Neutral atoms are a rapidly advancing platform for scalable quantum computing~\cite{Bluvstein_2024_neutralatomQC,baranes2025leveraging, chiu2025continuous, bluvstein2025architectural}. Each node consists of a large array of individually trapped atoms (e.g., Rubidium or Ytterbium), partitioned into computational and communication zones~\cite{baranes2025designing,pattison_fast_2024}. Local logic is performed using laser-driven Rydberg interactions, which enable high-fidelity entangling gates across many qubits. While the gate operations are fast, atomic motion currently limits overall local gate times to $\sim$200 $\mu$s~\cite{Bluvstein_2024_neutralatomQC}. Communication qubits couple to photonic modes to enable inter-node entanglement generation. Future architectures are expected to support scalable communication with high-rate photonic links and flexible memory allocations~\cite{sinclair_2025neutral_atoms_interconnect,Covey_neutral_atoms_quantum_networks}. These trade-offs are explored in detail in \cref{app:atoms landscape}, where we analyse performance across several photonic interconnect designs with neutral atoms.

\begin{table}[H]
    \scriptsize
    \centering
    \begin{tabular}{|l|l|l|}
    \hline
    \textbf{Parameter} & \textbf{Demonstrated} & \textbf{Future Goal} \\
    \hline
    Local gate error & $\sim$0.5\% \cite{Evered_high_fidelity_gate_atoms} & 0.1\% \\
    Local gate time & 200~$\mu$s \cite{Bluvstein_2024_neutralatomQC} & 10 $\mu$s \\
    Distr. gate error & 2\% \cite{ritter2012elementary}  & <1\% \cite{sinclair_2025neutral_atoms_interconnect} \\
    Distr. gate rate & 30~Hz \cite{ritter2012elementary} & 10~kHz -- 100~MHz \cite{sinclair_2025neutral_atoms_interconnect} \\
    Idling error rate & $10^{-5}$ per $\mu$s \cite{Bluvstein_2024_neutralatomQC} & $10^{-6}$ per $\mu$s \\
    Memory size & 6\,000 \cite{manetsch2024tweezer} & 10\,000-100\,000 \\
    \hline
    \end{tabular}
    \caption{Neutral atom platform parameters. Communication qubits (atoms in cavities) distribute entanglement between nodes, while computational qubits in the array perform local logic. Advanced cooling and trapping techniques give very large array sizes and long-lived qubit states, supporting a scalable modular architecture.}
\end{table}

\section{Resource-aware performance landscape for neutral atoms} \label{app:atoms landscape}

\begin{figure*}
    \centering
    \includegraphics[width=\textwidth]{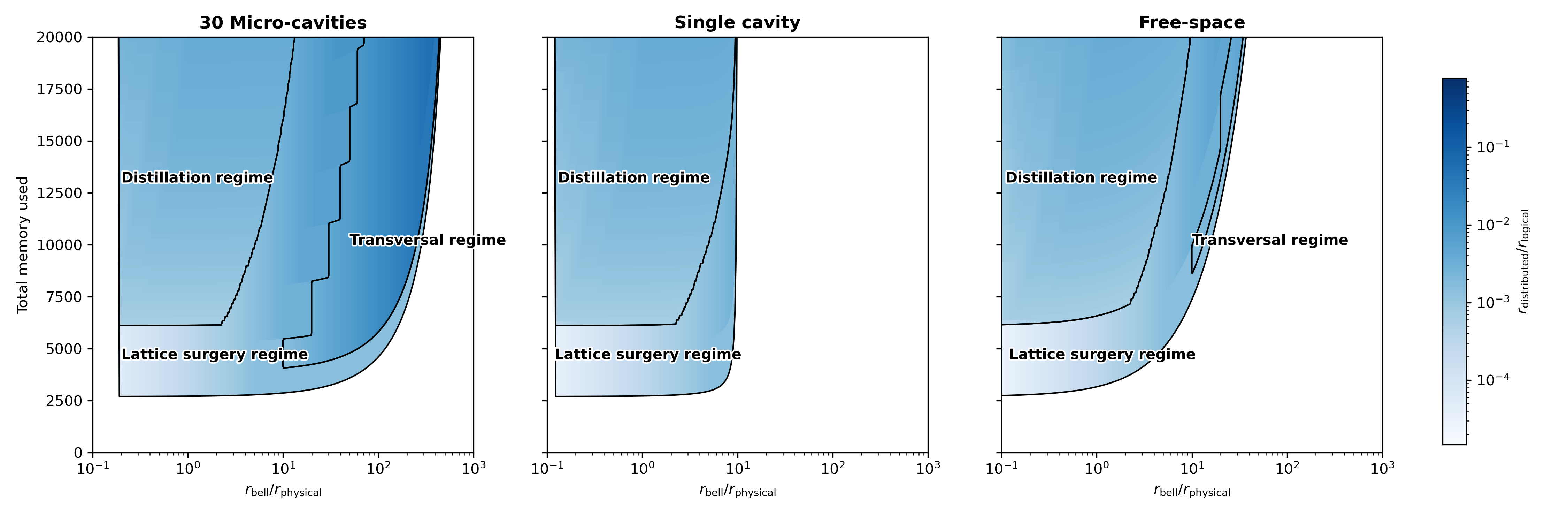}
    \caption{
    Simulated resource-performance map for distributed quantum computing with neutral atoms across three interconnect architectures.
    The x-axis shows the normalised Bell-pair rate $r_{\mathrm{bell}} / r_{\mathrm{physical}}$, and the y-axis represents the \emph{total physical memory} required (communication + logic).
    Colour intensity reflects the achievable logical rate ratio $r_{\mathrm{distributed logical}} / r_{\mathrm{local logical}}$, with distinct regimes highlighting the best quantum error correction method out of logical distillation, lattice surgery, and transversal gates.
    Each panel corresponds to a different photonic interface: 30 micro-cavities, a single shared cavity, and free-space emission.
    Feasible regions (above 2,500 logic qubits and below 20,000 total memory) are calculated using the analytical model from Ref.~\cite{sinclair_2025neutral_atoms_interconnect} based on cavity count, $P_{\mathrm{aa}}$, and $t_{\mathrm{base}}$, as described in the text.
    }
    \label{fig:neutral_atoms_regimes}
\end{figure*}

The generation rate of remote Bell pairs in neutral atom systems depends critically on the architecture of the photonic interface, as well as the number of atoms allocated as communication qubits. We consider three prominent designs, each representing a distinct trade-off between parallelism, complexity, and achievable entanglement rates:

\begin{itemize}
    \item \textbf{30 Micro-cavities:}  
    Each communication atom is individually coupled to a dedicated optical cavity, enabling highly parallel photon generation. This design maximises the Bell pair rate per communication qubit and is projected to achieve rates up to $10^8$ Hz with large arrays~\cite{sinclair_2025neutral_atoms_interconnect}.  
    The overhead, however, includes the physical integration of many cavities and the associated optical alignment.

    \item \textbf{Single cavity:}  
    Here, multiple communication atoms interact sequentially with a single cavity mode, enabling time-multiplexed entanglement generation~\cite{Li_2025neutral_atoms_interconnect}.  
    This approach reduces optical hardware requirements but introduces a trade-off: the Bell pair rate is limited by cavity cycle time, scaling less favourably with the number of communication qubits.

    \item \textbf{Free-space:}  
    In this architecture, atoms emit photons without the use of cavities, relying on collective emission or frequency encoding schemes~\cite{Covey_neutral_atoms_quantum_networks}.  
    While simpler to scale in hardware, the success probability per attempt is lower, leading to smaller achievable Bell rates for a given communication memory size.
\end{itemize}

We model the Bell-pair generation rate $r_{\mathrm{bell}}(N)$ as a function of the number of communication qubits $N$ using the analytical model reported in Ref.~\cite{sinclair_2025neutral_atoms_interconnect}:
\begin{equation}
    r_{\mathrm{bell}}(N) = \frac{P_{\mathrm{aa}}}{t_{\mathrm{base}} + \frac{16}{N} + \frac{100 P_{\mathrm{aa}}}{N}} \times 10^6,
\end{equation}
where $P_{\mathrm{aa}}$ is the per-attempt success probability and $t_{\mathrm{base}}$ is the base attempt time.  
We invert this relationship to determine the required communication memory $N$ for achieving a desired Bell pair rate $r_{\mathrm{bell}} = r_{\mathrm{bell}}$. The total memory required for a given operating point is then:
\[
\text{total memory} = \text{logic qubits} + N.
\]

\Cref{fig:neutral_atoms_regimes} shows the \textit{resource-aware performance landscape} for distributed quantum computing with neutral atoms across these three architectures. The x-axis denotes the normalised Bell-pair generation rate $r_{\mathrm{bell}} / r_{\mathrm{physical}}$, while the y-axis represents the total physical memory (computation + communication).

The colour scale indicates the distributed-to-logical gate rate ratio $r_{\mathrm{distributed}} / r_{\mathrm{logical}}$, with contours highlighting the optimal QEC regime (distillation, lattice surgery, or transversal gates) for each operating point.  

We assume a total available qubit budget of 20,000 physical atoms, with a minimum of 2,500 allocated to logical computation. Feasibility regions for each interconnect architecture are derived by mapping achievable $r_{\mathrm{bell}}$ values for different $N$, based on the parameters:
\begin{itemize}
    \item \textbf{30 Micro-cavities:} $P_{\mathrm{aa}} = 0.24$, $t_{\mathrm{base}} = 0.003\,\mu$s.
    \item \textbf{Single cavity:} $P_{\mathrm{aa}} = 0.10$, $t_{\mathrm{base}} = 0.1\,\mu$s.
    \item \textbf{Free-space:} $P_{\mathrm{aa}} = 0.0035$, $t_{\mathrm{base}} = 0.0\,\mu$s.
\end{itemize}

This approach reveals how memory allocation for communication qubits directly affects the optimal QEC strategy and the achievable distributed logical gate rate. The resulting performance map, shown in \cref{fig:neutral_atoms_regimes}, thus provides a comprehensive comparison of these architectures under realistic hardware constraints.

\end{document}